\documentclass[a4paper,fleqn,usenatbib]{mnras}

\usepackage{newtxtext,newtxmath}

\usepackage[T1]{fontenc}
\usepackage{ae,aecompl}

\usepackage{graphicx}	
\usepackage{amsmath}	
\usepackage{amssymb}	

\newcommand{\beq}{\begin{equation}}
\newcommand{\beqa}{\begin{eqnarray}}
\newcommand{\eeq}{\end{equation}}   
\newcommand{\eeqa}{\end{eqnarray}}  

\title[A magnetized thin accretion disk]{A magnetized thin accretion disk:
numerical simulations compared with asymptotic expansion}

\author[M. \v{C}emelji\'{c}, W. Klu\'{z}niak \& V. Parthasarathy]{
Miljenko \v{C}emelji\'{c}$^{1,2}$\thanks{E-mail: miki@camk.edu.pl},
W. Klu\'{z}niak$^{1}$ \& V. Parthasarathy$^{1,3}$\\
$^{1}$Nicolaus Copernicus Astronomical Center, Bartycka 18, 00-716
Warsaw, Poland\\
$^2$Academia Sinica Institute of Astronomy and Astrophysics, P.O. Box
23-141, Taipei 106, Taiwan\\
$^3$H\"ochstleistungsrechenzentrum Stuttgart, Nobelstra\ss e 19, 70569 Stuttgart,
Germany\\
}

\date{Accepted XXX. Received YYY; in original form ZZZ}

\pubyear{2019}

\begin{document}
\label{firstpage}
\pagerange{\pageref{firstpage}--\pageref{lastpage}}
\maketitle

\begin{abstract}
To obtain a simple description of a geometrically thin magnetic
accretion disk, we apply the method of asymptotic expansion. For the
first time we write a full set of stationary asymptotic approximation
equations of a thin magnetic accretion disk. As the obtained equations
cannot be solved without knowledge of the solutions at the disk surface,
we combine the results from numerical simulations and from analytical
equations to find a simple set of functional expressions describing the
radial and vertical dependence of physical quantities in the disk. Except
very close to the star, the functional form of the disk variables is quite
similar in the HD and MHD cases, with the overall scale of density,
and the vertical and radial velocity components modified by the stellar
magnetic field.
\end{abstract}

\begin{keywords}
Stars: formation, pre-main sequence, -- magnetic fields --MHD 
\end{keywords}


\section{Introduction}
The gravitational infall of matter onto a rotating central object
naturally leads to the formation a rotating accretion disk.  The
matter from the disk is fed inwards through the accretion column onto
the magnetized central star.  The first analytical solution for the
accretion disk flows has been given in \citet{ss73}, who also proposed
a prescription for the viscosity coefficient ($\alpha$ times the pressure).
In that
$\alpha$-disk model, and in many following works, the radial solution
was obtained as an average over the disk thickness, with equations in
the vertical direction separately solved to obtain a hydrostatic
balance.

\citet{urp84} has shown that the proper description of the accretion
flow cannot be obtained by its height-averaged values. This follows
from vertical gradients of the stress tensor, which cause the flow
direction in the midplane of the disk to be opposite to that in the
subsurface layers. This conclusion was upheld by the results in
numerical simulations.  A hydrodynamical (HD) solution of a steady
axisymmetric, polytropic accretion disk in three dimensions has been
given in \citet[hereafter KK00]{KK00}; the velocity field was found to
exhibit backflow in the equatorial regions for all values of the
viscosity parameter, $\alpha$, greater than a certain critical
value. The solution was extended numerically in \citet{Reg02} to the
ideal equation of state with radiative losses.

We generalize the KK00 solution to the case of a
magnetic disk. Since the solution inside the disk depends on details
of the star-disk magnetospheric interaction,
we cannot write
separate solutions in the disk without knowing the global solution.
From the obtained equations, only general conditions on the
magneto-hydrodynamic (MHD) solution can be given.

In previous work the induction equation
was solved assuming a prescibed velocity field in the disk, following
from the HD disk solutions, e.g, the \citet{ss73} in  \citet{naso1,naso2},
or the {KK00} solution in \citet{naso3}.
Here we allow the magnetic field to influence the flow,
in this sense we are self-consistently solving for the fluid velocity field.

To obtain magnetic solutions in numerical simulations, we
use the HD solution from KK00 disk as an initial condition,
adding a hydrostatic corona and the stellar magnetic field
between the rotating stellar surface and the accretion disk.
Results in our simulations were shown in  \cite{cem19}, where we obtained
quasi-stationary solutions with different stellar rotation rates,
magnetic field strengths, and magnetic Prandtl numbers. Here we
confront the disk solutions from such simulations with the
requirements obtained from the analytical equations. To do this,
we match the solutions in the disk with a set of expressions which
best describe our numerical simulations.

In the following, in \S 2 we present the equations which we are solving,
and outline the results of the method of asymptotic approximation
in \S 3. In \S 4 we present the quasi-stationary results
of our numerical simulations, and find the
expressions for the best matches to the numerical solutions. In \S 5
we compare the numerical solutions with the analytically obtained
conditions, discussing the changes in our results with the different
physical parameters in \S 6 and summarizing in \S 7. The Appendix
gives a detailed derivation of the equations in the method of
asymptotic approximation, and includes graphs of the
solutions and matching functions in our numerical simulations.

\section{Resistive accretion disk in stellar magnetic field}\label{sec1}
\begin{figure}
\includegraphics[width=\columnwidth]{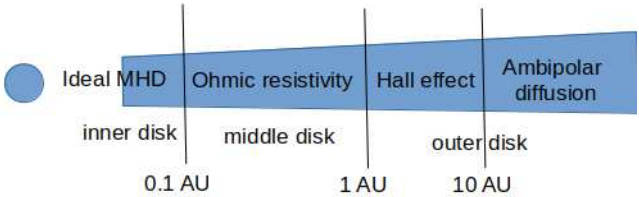}
\caption{Illustration of the reach of the inner, middle and outer disk
regions in the case of Young Stellar Objects. In the inner disk region
the disk is in the ideal MHD regime. The Ohmic resistivity adds to the
viscous dissipation in the middle disk region, and in the outer disk
other resistive terms prevail in the induction equation.
}
\label{fig:diskreg}
\end{figure}

Following KK00, where the equations for the viscous, hydrodynamical
case of the thin accretion disk were derived and solved in  a systematic,
term by term, expansion in the dimensionless thickness of the disk, we
derive the equations for the magnetized, resistive accretion disk.
In the asymptotic expansion we
consider also, for the first time, the energy equation.

In the MHD case the obtained equations cannot be solved
without knowledge of the solutions at the disk surface,
and these, in turn, depend on the magnetic field interaction
with the star through the star-disk magnetosphere.
For this reason a numerical solution of the equations is necessary.
Nonetheless, useful constraints can be obtained from the asymptotic
expansion equations, and they suggest a functional form for various
physical variables, allowing us to extract the radial and vertical
dependence of the variables from the results of numerical
simulations.

 We are solving the viscous and resistive
equations of magneto-hydrodynamics which are, in the cgs system of units:
\beqa
\frac{\partial\rho}{\partial t}+\nabla\cdot(\rho\mathbf{v}) =0\\
\nabla\cdot\mathbf{B}=0\\
\frac{\partial\rho\mathbf{v}}{\partial t}
+\nabla\cdot\left[\rho\mathbf{v}\mathbf{v}
+\left(P+\frac{B^2}{8\pi}\right)\tilde{I}-
  \frac{\mathbf{B}\mathbf{B}}{4\pi}-\tilde{\tau}\right]=\rho\mathbf{g}\\
\frac{\partial E}{\partial t}
+\nabla\cdot\left[\left(E+P+\frac{B^2}{8\pi}\right)\mathbf{v}-\frac{(\mathbf{v}
\cdot\mathbf{B})\mathbf{B}}{4\pi}\right]=\rho\mathbf{g}\cdot\mathbf{v}\\
\frac{\partial\mathbf{B}}{\partial t}+\nabla\times(\mathbf{B}
\times\mathbf{v}+\eta_{\mathrm m}\mathbf{J})=0, 
\label{eqsss}
\eeqa
where $\rho$, $P$, $\mathbf{v}$, $\mathbf{B}$ and  $\eta_{\rm m}$ are the density, pressure,
velocity, magnetic field and the Ohmic resistivity, respectively. The terms
$\tilde{I}$ and $\tilde{\tau}$ are representing the unit tensor and the viscous stress
tensor, respectively.

We search for the quasi-stationary state solutions, assuming that all the
heating is radiated away from the disk. For this reason, the dissipative
viscous and resistive terms are not present in the energy equation. We still
solve the equations in the non-ideal MHD regime, because of the viscous term
in the momentum equation, and the Ohmic resistive term in the induction equation.

In our numerical simulations, we consider a part of the star-disk system
close to a central object, with the physical domain reaching only into the
middle region of the disk, shown in Fig.~\ref{fig:diskreg}, where only
the Ohmic dissipation takes place,
in addition to the viscous one. In the cases of different objects,
those regions reach different physical distances. Here we present the
case of Young Stellar Objects (YSOs), in which our disk reaches to
$R_{\rm max}<0.5$~AU. Inside 0.1 AU from the Young Stellar Object,
the disk is in the ideal MHD regime, with the ``frozen in'' magnetic field
inside the star-disk magnetosphere. To the distance of about 1~AU from the
central YSO, the Ohmic resistivity is the largest contributor to
the dissipation in the induction equation. Further away, the Hall
resistive term becomes most important, and even further away from the
central object, the ambipolar diffusion prevails.

The acceleration of gravity  is $\mathbf{g}=-\nabla\Phi_{\mathrm g}$, and
the gravitational potential of the star with mass $M_\star$ is equal to
$\Phi_{\mathrm g}=-GM_\star/R$. The total energy density
$E=P/(\gamma-1)+\rho\varv^2/2 + B^2/(8\pi)$
and the electric current is given by the Ampere's law
$\mathbf{J}=\nabla\times\mathbf{B}/(4\pi)$.
We assume the ideal gas with the
plasma adiabatic index $\gamma=5/3$, corresponding to polytropic index
$n=3/2$.

To compare the magnitude of the different terms in the equations
the equations have to be written in normalized units.
The comparison allows a partial solution and this is done in the Appendix. 
In the following Section we summarize the results of this analytic approach.

\section{Thin magnetic accretion disk in asymptotic approximation}
\label{asec1}
In the asymptotic approximation, pioneered in the context of accretion
disk by \cite{Reg83}, all the variables are written in the Taylor
expansion, with the coefficient of expansion
given by the characteristic ratio of disk height to the radius,
$\epsilon=\tilde{H}/\tilde{R}<<1$ (see KK00, and also \cite{um06} for a
general discussion of the asymptotic approximation. We can compare the
terms of the same order in $\epsilon$ for each variable X, and then
write the result as
$X=X_0+\epsilon X_1+\epsilon^2 X_2+\epsilon^3 X_3+\dots $.

In the case of a viscous HD disk ($\mathbf{B}=0$),
equations of the previous section
could be solved inside the disk (KK00).
When a stellar magnetic field is
present, a solution in the disk cannot be separated from the star-disk
magnetosphere, because of the connection of the magnetic field in the
corona with the field in the disk. In addition, the solution in the
magnetosphere can itself be complicated by the reconnection events and
outflows, and a back-reaction from the disk.

For the reader's convenience, the Appendix gives a step-by-step
example of the asymptotic approximation in the equation of continuity,
together with a condensed derivation
of the conditions that the solutions obtained
from the zeroth, first and second order in $\epsilon$ should satisfy
in the complete set of the viscous and resisitive MHD equations.

We give an outline of the results, obtained by the method of
asymptotic approximation. In the magnetic case we can only obtain a
general set of conditions that should be satisfied in a self-consistent
solution.
Later in the text we check if the results of our numerical simulations
satisfy those conditions.

From the radial component of the momentum equation we readily obtain
$\Omega_0=1/r^{3/2}$. This solution is valid equally in the HD and MHD
cases.

As seen in the Appendix, the zeroth order magnetic field in the disk is
a function of the radius alone. If we insert the $B_0=f(r)$ condition
into the vertical component of the momentum equation in the zeroth order
in $\epsilon$ (Eq.~\ref{vertmom0}), we obtain the vertical hydrostatic
equilibrium condition. It gives the same
solution for the lowest order in $\epsilon$ for the density (see
Eq.~\ref{rhozero}), pressure and the sound speed as were obtained in the
purely HD solution. The difference from KK00 is that now the disk
surface boundary condition is not vacuum, but a corona with the density
$\rho_{\rm cd}(r)$ at the disk interface.
The zeroth order profile of density, pressure, and the sound speed are
\beq
\rho_0(r,z)=\left[\rho_{\rm cd}^{2/3}(r)+\frac{h^2-z^2}{5r^3}\right]^{3/2},
\nonumber
\eeq
\beq
P_0(r,z)=\left[\rho_{\rm cd}^{2/3}(r)+\frac{h^2-z^2}{5r^3}\right]^{5/2}, \nonumber
\eeq
\beq
c_{\rm s{0}}(r,z)=
\sqrt{\frac{5}{3}\left[\rho_{\rm cd}^{2/3}(r)+\frac{h^2-z^2}{5r^3}\right]}.
\label{sound}
\eeq
Clearly, $\rho_0(r,h)=\rho_{\rm cd}(r).$

Far away from the star, where we expect a small effect of the magnetic
field, solutions in the simulations should not differ much from the HD
solutions. Closer to the star, the magnetic field influence increases and
the change in results will be larger. Higher order terms in the MHD
solution may differ from those of KK00.

We now list the main conditions on the solution obtained in the Appendix.
Expanding the stationary and axi-symmetric normalized analytical equations
in the small parameter $\epsilon=H/R$, with the assumed vertical symmetry
accross the disk equatorial plane, we find:\\
$\bullet$ $\varv_{r0}=\varv_{z0}=\varv_{z1}=\Omega_1=c_{\mathrm s1}=\rho_1=B_{\mathrm r0}=0$,\\
$\bullet$ The quantities $B_{\mathrm z0}$, $B_{\varphi 0}$, $B_{\mathrm z1}$,
$B_{\varphi 1}$, $B_0$, $B_1$ are all $f(r)$ only,\\
$\bullet$ $\Omega_0=r^{-3/2}$, $B_{\mathrm z2}=zf(r)$,
$\partial_z c_{\mathrm s0}^2 =-z/(nr^3)$.

\section{Results from numerical simulations}
\begin{figure}
\includegraphics[width=\columnwidth]{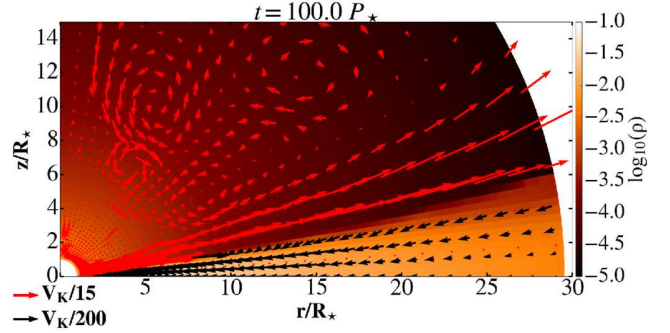}
\caption{Capture of our hydrodynamic simulation after t=100
stellar rotations. The matter density is shown in logarithmic
color grading in code units, with a sample of velocity vectors.
Since the poloidal velocity in the corona is much larger than in the
disk, velocity vectors are shown with a different scaling, as indicated
by the arrows below the panel corresponding to multiples of the
Keplerian velocity at the stellar surface.}
\label{fig:hdsol}
\end{figure}
\begin{figure}
\includegraphics[width=\columnwidth]{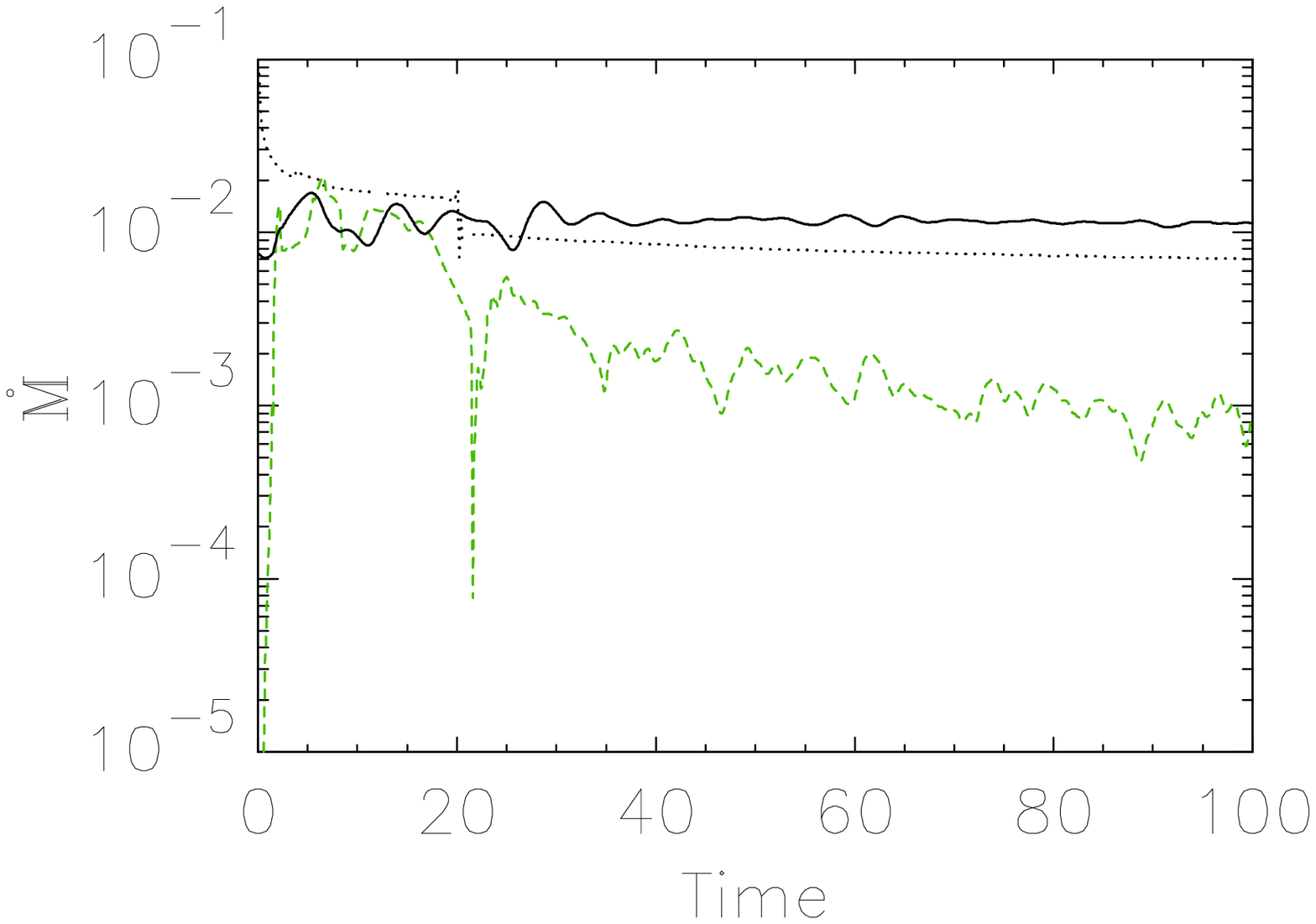}
\includegraphics[width=\columnwidth]{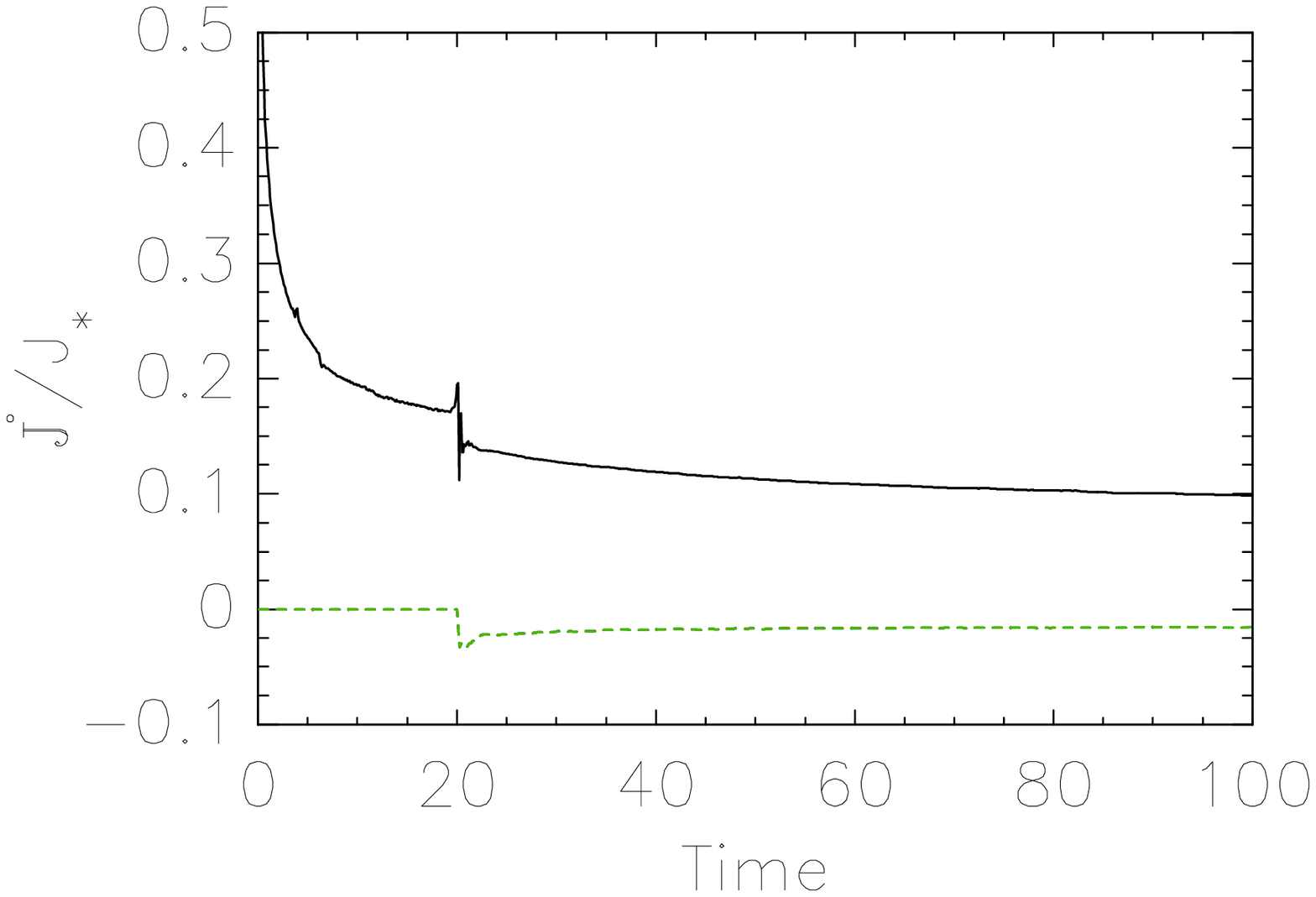}
\caption{Illustration of the quasi-stationarity of our solution in the
HD case. {\it Top panel:}  evolution in time of the mass flux in the
wind, onto the star and through the disk at r=15$R_\star$ (dashed green,
dotted and solid lines, respectively), in units of
$\dot{M}_0=\rho_{\rm d0}R_\star^3\Omega_{\mathrm K\star}$.
{\it Bottom panel:} the torque exerted  on the stellar surface by
the matter accreted from the disk, and by the wind (solid and dashed
green lines, respectively) in units of
$\dot{J}_0=\rho_{\rm d0}R_\star^5\Omega_{\rm K\star}^2$ per stellar
angular momentum $J_\star=k^2M_\star R_\star^2\Omega_\star$. For the
typical normalized gyration radius of a fully convective star we use
$k^2=0.2$. Positive torque spins-up, and negative torque spins-down the
star. Time is measured in the number of stellar rotations.}
\label{fig:maccmominthd}
\end{figure}
\begin{figure}
\includegraphics[width=\columnwidth]{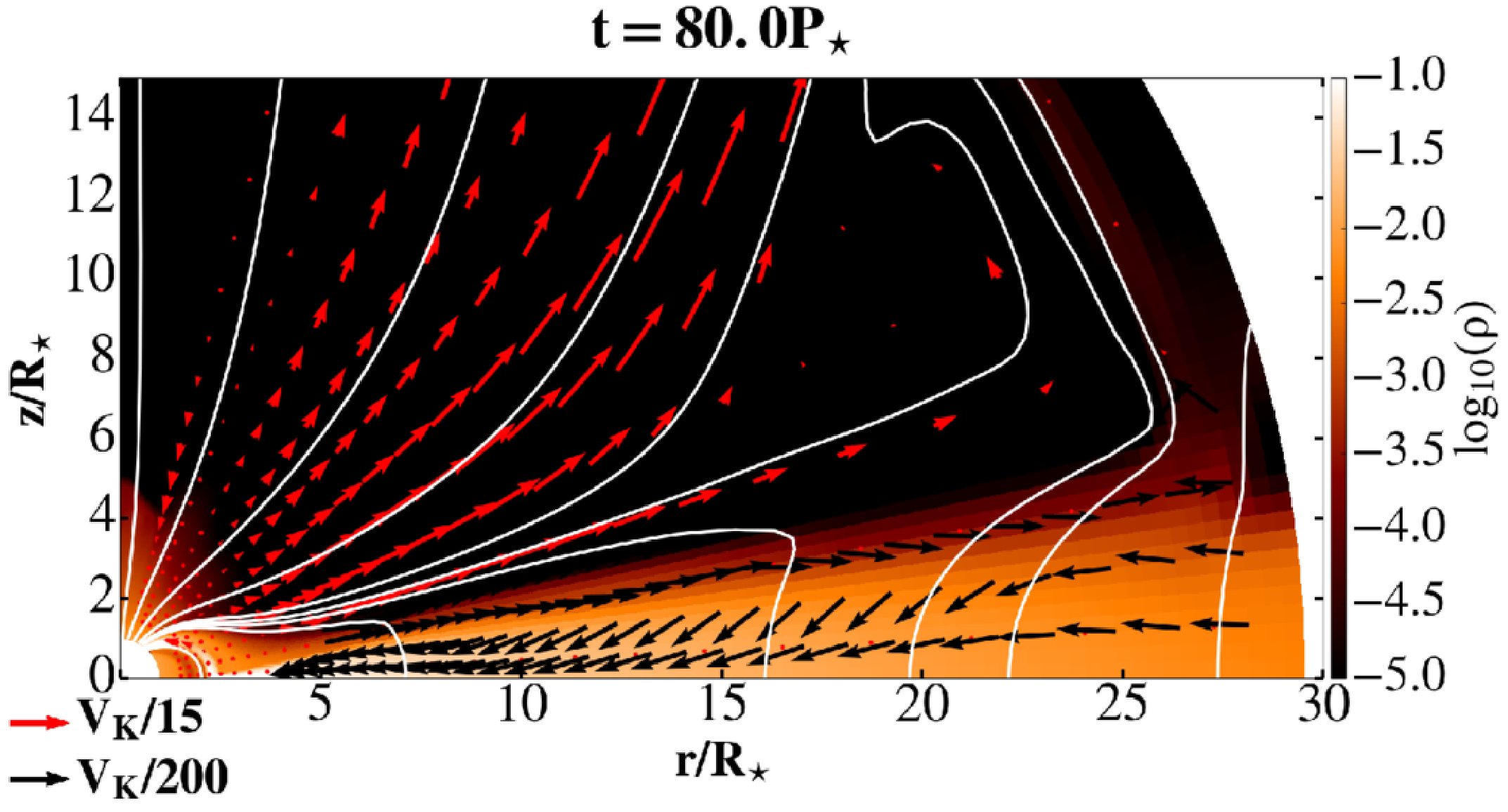}
\includegraphics[width=\columnwidth]{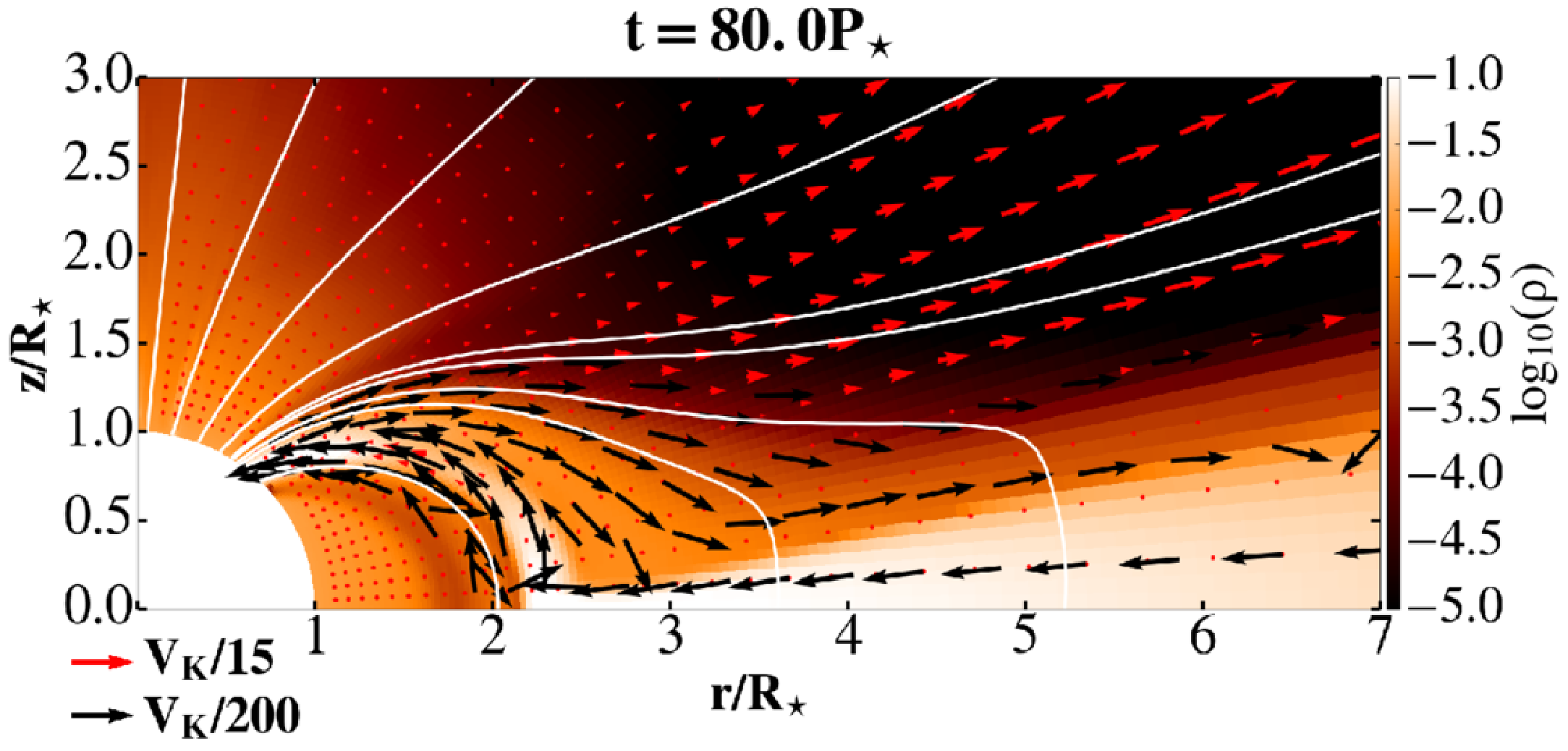}
\caption{Captures of our magnetic simulation after t=80 stellar
rotations (top panel), and a zoom closer to the star (bottom panel)
to better show the accretion column. Colors and vectors have the same
meaning as in Fig.~\ref{fig:hdsol}.
Note the different scale of the poloidal velocity (arrows below the panels).
A sample of the poloidal magnetic field lines is shown with the solid lines.
}
\label{fig:mhdsol1}
\end{figure}
\begin{figure}
\includegraphics[width=\columnwidth]{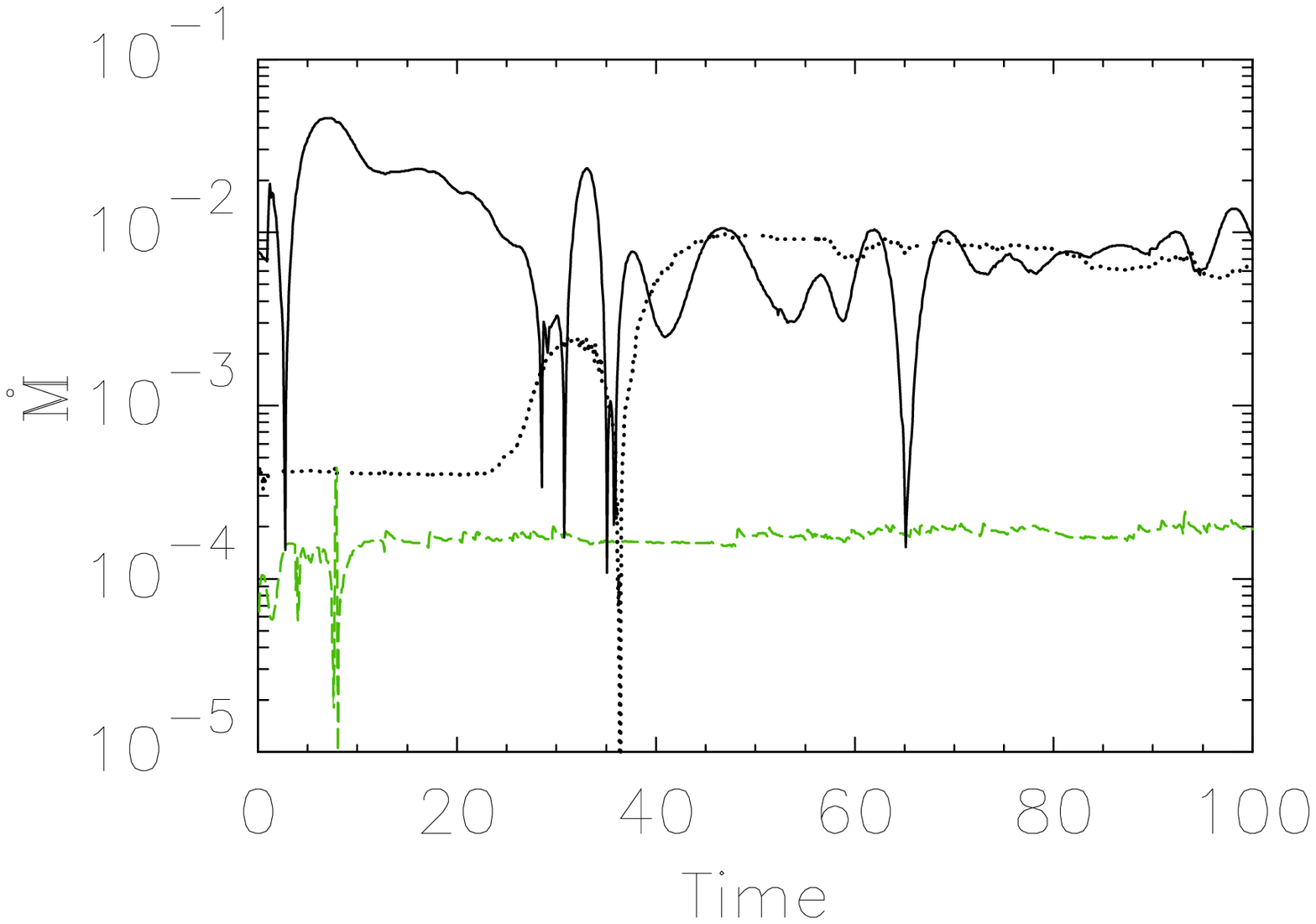}
\includegraphics[width=\columnwidth]{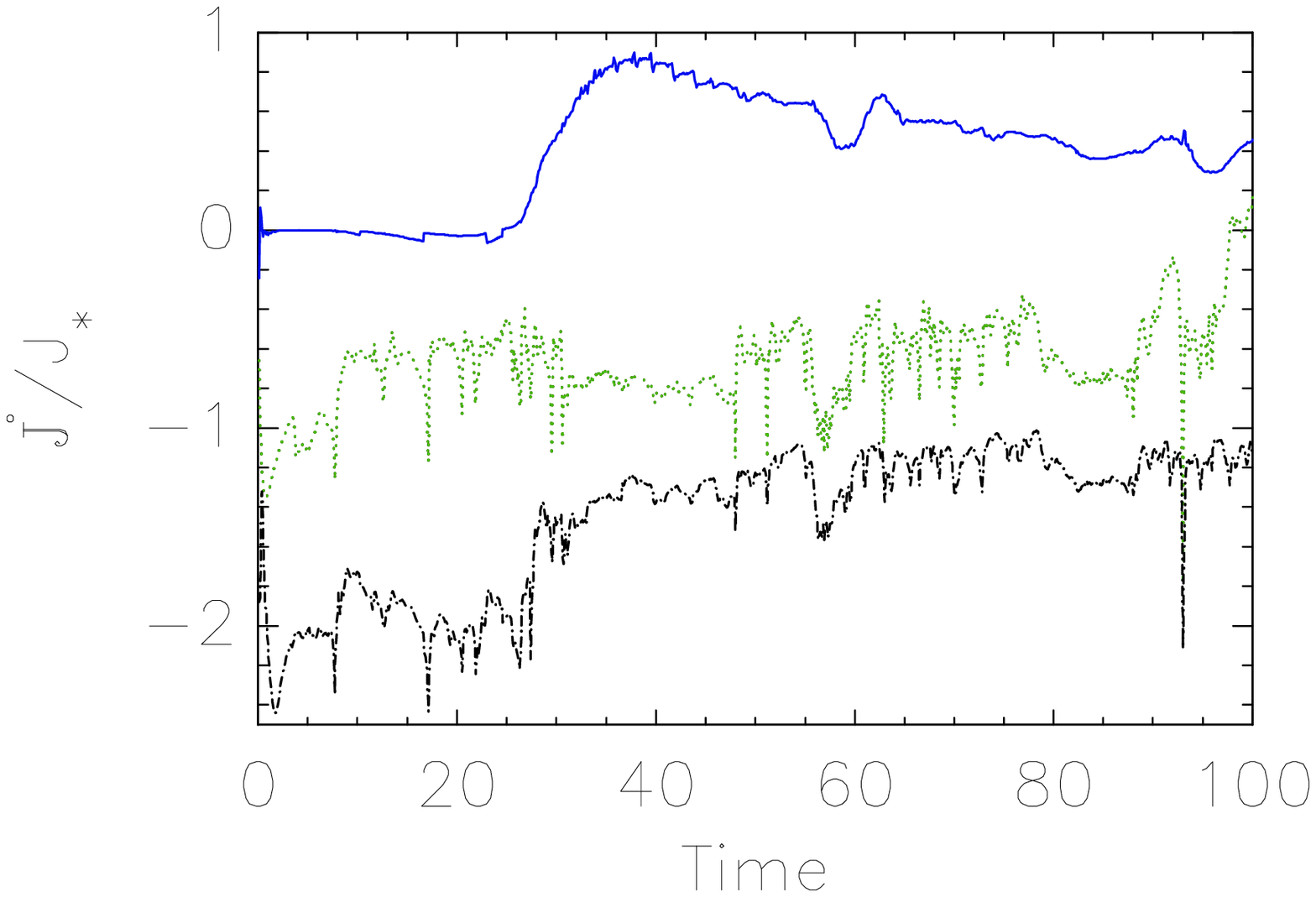}
\caption{Illustration of the quasi-stationarity in our magnetic case
solutions, in the same units as in Fig.~\ref{fig:maccmominthd}.
{\it Top panel:} the mass flux in the various components of
the flow through the disk at r=15$R_\star$ (solid line), onto the stellar
surface (dot-dashed line) and into the stellar
wind (dashed green line). {\it Bottom panel:} the torques in the
different components of the flow in the wind (dotted green line), in the
matter falling onto the star from the part of the disk beyond
$R_{\mathrm{cor}}$ (dot-dashed line) and below $R_{\mathrm{cor}}$
(solid blue line).}
\label{fig:maccmomint1}
\end{figure}

Extensive umerical simulations with a KK00 disk as an initial condition
were performed in \cite{cem19}, following \cite{zf09}. Here we give a brief
overview of the setup. We solve the non-ideal MHD equations using the {\sc pluto}
(v.4.1) code \citep{m07,m12} in the spherical grid. The resolution is
$R\times\theta=(217\times 100)$ grid cells, in a logarithmically
stretched radial grid and in a half of the meridional half-plane in a
uniform grid $\theta=[0,\pi/2]$. The viscosity and resistivity are
parameterized by the \citet{ss73} $\alpha$-prescription as proportional
to $c_{\mathrm s}^2/\Omega_{\mathrm K}$. For the magnetic field, a
split-field method is used, so that we evolve in
time only changes from the initial stellar magnetic field
\citep{tan94,pow99}, with the constrained transport method used to maintain
the $\nabla\cdot{\mathbf B}=0$. Simulations were performed using the
second-order piecewise linear reconstruction and an approximate Roe solver.
The second-order time-stepping (RK2) was employed.

Here we present the results in our HD and non-ideal MHD numerical simulations
of a YSO, in the physical domain reaching 30 stellar radii,
$R_{\mathrm{max}}=30R_\star$, with the
(anomalous\footnote{Anomalous diffusive coefficients are much larger
their microscopic equivalent. They are usually given as free parameters in the
simulations, assuming that dissipation is a result of turbulence.}) viscosity
parameter $\alpha_{\rm v}=1$ and the mass
accretion rate in the disk $\dot{M}_0=5\times10^{-7}~M_\odot/{\rm yr}$.
The stellar rotation rate is taken to be
$\Omega_\star=0.2~\Omega_{\mathrm{br}}$, where $\Omega_{\mathrm{br}}$ is
the equatorial mass-shedding limit rotation rate, equal to the
Keplerian angular velocity for the star
$\Omega_{\mathrm K\star}=\sqrt{GM_\star/R_\star^3}$. Thus, the
corotation radius is
$R_{\rm{cor}}=(GM_\star/\Omega_\star^2)^{1/3}=(0.2)^{-2/3}R_\star\approx2.9R_\star$.
In the Classical T-Tauri star case, the stellar mass is
$M_\star=0.5M_\odot$, radius
$R_\star=2R_\odot$, the Keplerian velocity at the stellar equator is
$\varv_{\mathrm K\star}=218~km/s$ and the stellar rotation period is
$P_\star=2\pi/\Omega_\star=2.3$ days. Then
$\rho_{\mathrm d0}=1.2\times10^{-10}g/cm^3$. In the magnetic case we add the
stellar dipole field of $B_\star=500~G$, and the resistivity parameter
$\alpha_{\rm m}=1$, so that the magnetic Prandtl number
$P_{\rm m}=2\alpha_{\rm v}/(3\alpha_{\rm m})=0.67$.

A table for rescaling to other types of objects is given in \cite{cem19}
where we performed a parameter study with the same set-up. We varied
the stellar rotation rate, magnetic field strength and resistivity in the
disk and compared the changes in results in dependence on those parameters.

We output the results along the $z$ axis at two radial positions in the disk:
in the middle of the radial domain, which lies far behind the distance $r_m$,
where the viscous torque is vanishing\footnote{The distance $r_m$ defines a
natural length scale $r_+=\Omega_m^2r_m^4/(GM_\star)$, with $\Omega_m$ the
Keplerian rotation rate at $r_m$, see KK00. The outer region of the disk is
at a much larger radius.} and closer to the star, just behind the corotation
radius. We derive two sets of expressions along the vertical direction
from those results, one at each distance from the star. Along the spherical
radial direction, we output the results in the disk along a line near to the
disk equator, and also along a line near to the disk surface. For each
physical quantity, we verify if there is a unique solution throughout the
disk.

Starting from the analytical solution as an initial condition in the
simulations, we obtain a numerical solution. We then compare
the quasi-stationary solutions in both the HD and the MHD solution, to
the initial condition (i.e. the analytical solution) itself. The
quasi-stationary solution does not change much in the final
several tens of stellar rotations in our simulations.
The magnetic field and the accretion rate of the observed stars are
practically constant during such an interval, so that our
time-independent analytical solutions are a good representation of
the solutions.

Our computational domain reaches into the middle disk region, shown in
Fig.~\ref{fig:diskreg}, where the resistivity adds to the viscosity as a
dissipation mechanism. This could make some of the assumptions from the
purely HD disk implausible---we check whether or not this is true
with the help of numerical simulations. We find that the magnetic
solutions follow the HD solutions in the functional dependence, only
the proportionality constants change.

A capture of our HD solution after 100 stellar rotations is shown in
Fig.~\ref{fig:hdsol}.
The poloidal fluid velocity vectors are represented by the arrows,
red for the corona, black for the disk, with a different scaling
(one unit of arrow length corresponding to velocities in the corona and the
disk in the ratio 40:3).
In this case, accretion onto the star proceeds
through the disk connected to the stellar equator. The mass and angular
momentum fluxes onto the star and into the wind during the simulation are
shown in Fig.~\ref{fig:maccmominthd}. 

The solution in the magnetic case is shown in Fig.~\ref{fig:mhdsol1}. When the
stellar dipole field is large enough, an accretion column is formed from the
inner disk rim onto the stellar surface near the polar region.
The matter is lifted above the disk
equatorial plane, following the magnetic field lines. The mass flux onto the
star and into the wind is shown in Fig.~\ref{fig:maccmomint1},
together with the angular momentum fluxes, shown in the second panel in the
same figure.

To investigate how much the magnetic solutions depart from the HD ones, and
from the KK00 analytical solution, we directly compare the density and
velocity profiles. Since the KK00 solutions are obtained in the cylindrical
coordinates, which are more convenient to plot, we project our results from
the simulations in spherical coordinates to the cylindrical coordinates.
The results are listed in the Appendix. In all the cases we also show the
closest match\footnote{Our approximate matches are not formal fits, but
the simplest functions following the quasi-stationary solution. In most
cases when the solution is without oscillations, the match is inside the
10\% of the solution. As illustrated in the Appendix, if oscillations
are present, the error can be larger.} to the case with
$B_\star=500$ G. 

We can write the results in our simulations as simple functions with
coefficients of proportionality:
\beqa\label{compeqsa}
 \rho(r,z)=\frac{k_1}{r^{3/2}}\left[1-\left(\zeta_1\frac{z}{r}\right)^2\right],
\eeqa
\beqa
 \varv_r(r,z)=\frac{k_2}{r^{3/2}}\left[1+\left(\zeta_2 z\right)^2\right],  \\ \nonumber
 \varv_z(r,z)=\frac{k_3}{r}z^{3/2},  \\ \nonumber
 \varv_{\varphi}(r,z)=\frac{k_4}{\sqrt{r}},\
 \Omega=\frac{\varv_{\varphi}}{r}=\frac{k_4}{r^{3/2}}.   \nonumber 
\eeqa

The momentum in the (cylindrical) radial direction can be written as:
\beq
\rho\varv_r(r,z)=\frac{k_1k_2}{r^3}\left[1-\left(\zeta_1\frac{z}{r}\right)^2\right]\left[1+\left(\zeta_2z\right)^2\right].
\eeq
Magnetic field components are proportional to $r^{-3}$, as expected for the
dipole stellar field, and depend linearly on height above the disk midplane:
\beq
B_r(r,z)=\frac{k_5}{r^3}z,\ B_z(r,z)=\frac{k_6}{r^3}z,\
B_\varphi(r,z)=\frac{k_7}{r^3}z, 
\label{compeqsb}
\eeq
In the case of $B_{r}$, the linear dependence is a consequence of the
boundary condition at the disk equatorial plane, where the magnetic field
components are reflected, with the change in sign of the component
tangential to the boundary. This means that the radial magnetic field
component $B_{r}\rightarrow 0$  at the equatorial plane, and is slowly
increasing above (and below) that plane, in the densest parts of the disk.
It is catching-up with more dramatic changes only close to the disk maximal
height at the given radius, where it matches the values in the corona above
the disk.

The vertical dependence of the viscous and resistive dissipative
coefficients $\eta$ and $\eta_{\rm m}$ in the initial conditions was
taken to follow the $(z/r)^2$ dependence of $c_{\rm s0}^2$ from
Eq.~(\ref{sound}) in the analytical solution in Eq.~\ref{pressound},
which can be further written as in Eq.~\ref{cskvadr}. The same
dependence is found in the results of our simulations, in both inner
and outer parts of the disk:
\beq
\eta(r,z)=\frac{k_8}{r}\left[1-\left(\zeta_8 \frac{z}{r}\right)^2\right],
\quad\eta_{\mathrm m}(r,z)= k_9\sqrt{r}
\left[1-\left(\zeta_9\frac{z}{r}\right)^2\right]. 
\label{etasoba}
\eeq

We assign the proportionality coefficients as $k_1, k_2, \dots$ in
the cases with a stellar dipole field of 500~G and 1000~G in
Table~\ref{tabsols}, indicating by the additional subscripts $i$ and $o$
if they are given in the {\it inner} (R=6) or {\it outer} (R=15) position
in the disk\footnote{Not to be mixed with the inner and outer regions from
\S\ref{sec1}.}. We also assign the corresponding coefficients $\zeta_1,
\zeta_2, \dots$ where needed.

\begin{table}
\caption{The proportionality coefficients in our simulations with
B$_\star$=0.5~kG and 1~kG.}
\label{tabsols} 
\centering                          
\begin{tabular}{ c c c }        
\hline    
B(kG) &  0.5 &  1 \\
\hline  
coef. & R=6 | R=15 & R=6 | R=15  \\
\hline\hline
$k_{1}$ & 0.87 & 1.2 \\
$k_{2i}$|$k_{2o}$ & -0.066 | -0.087 & 0.001 | -0.036  \\
$k_{3i}$|$k_{3o}$ & $4\times10^{-5}$|$3.75\times10^{-5}$ & 
$5.5\times10^{-4}$|1.2$\times10^{-5}$  \\
$k_4$ & 0.255 & 0.255  \\
$k_{5i}$|$k_{5o}$ & -0.69 | -0.41 &  -1.25 \\
$k_{6i}$|$k_{6o}$ & -0.35 | -0.15 & -0.25 \\
$k_{7i}$|$k_{7o}$ & -2.59 | -1.13 & -7.99 | -1.2  \\ 
$k_8$ & 0.006 & 0.008  \\
$k_9$ & 0.01 & 0.01 \\
\hline
$\zeta_{1}$ & 6 & 6 \\
$\zeta_{2i}$|$\zeta_{2o}$ & 0.001 | 0.5 & 15.0 | 0.8 \\
$\zeta_{8}$ & 6.8 & 6.8 \\
$\zeta_{9}$ & 4.5 & 4.5 \\
\hline\hline     
\end{tabular}
\end{table}

In the following, we compare the above matches to solutions obtained
in the simulations, with the conditions obtained from the analytical
equations in the magnetic case.

\section{Comparison of the analytical and numerical solutions}
%
\begin{figure*}
\centering
\includegraphics[width=\columnwidth]{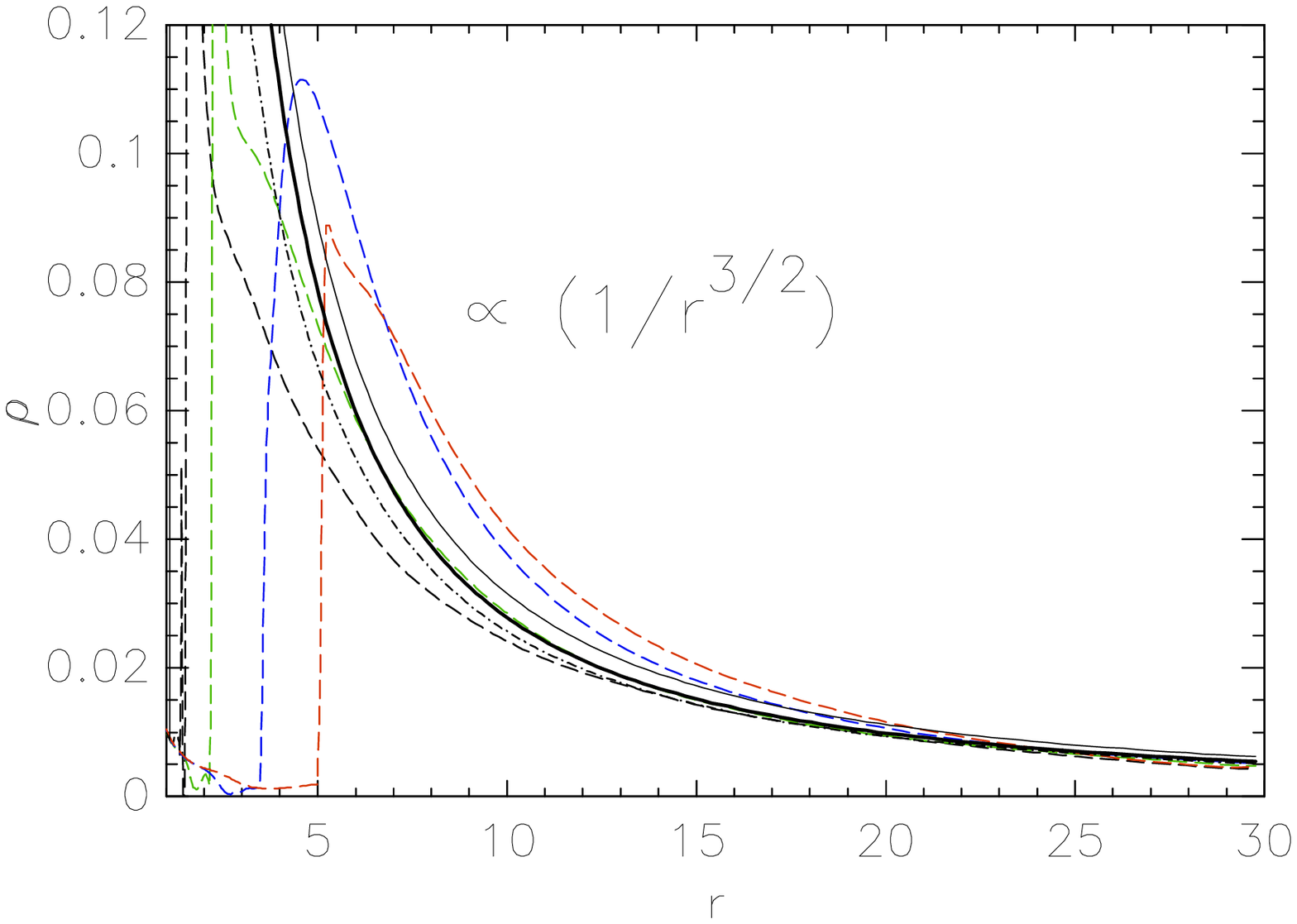}
\includegraphics[width=\columnwidth]{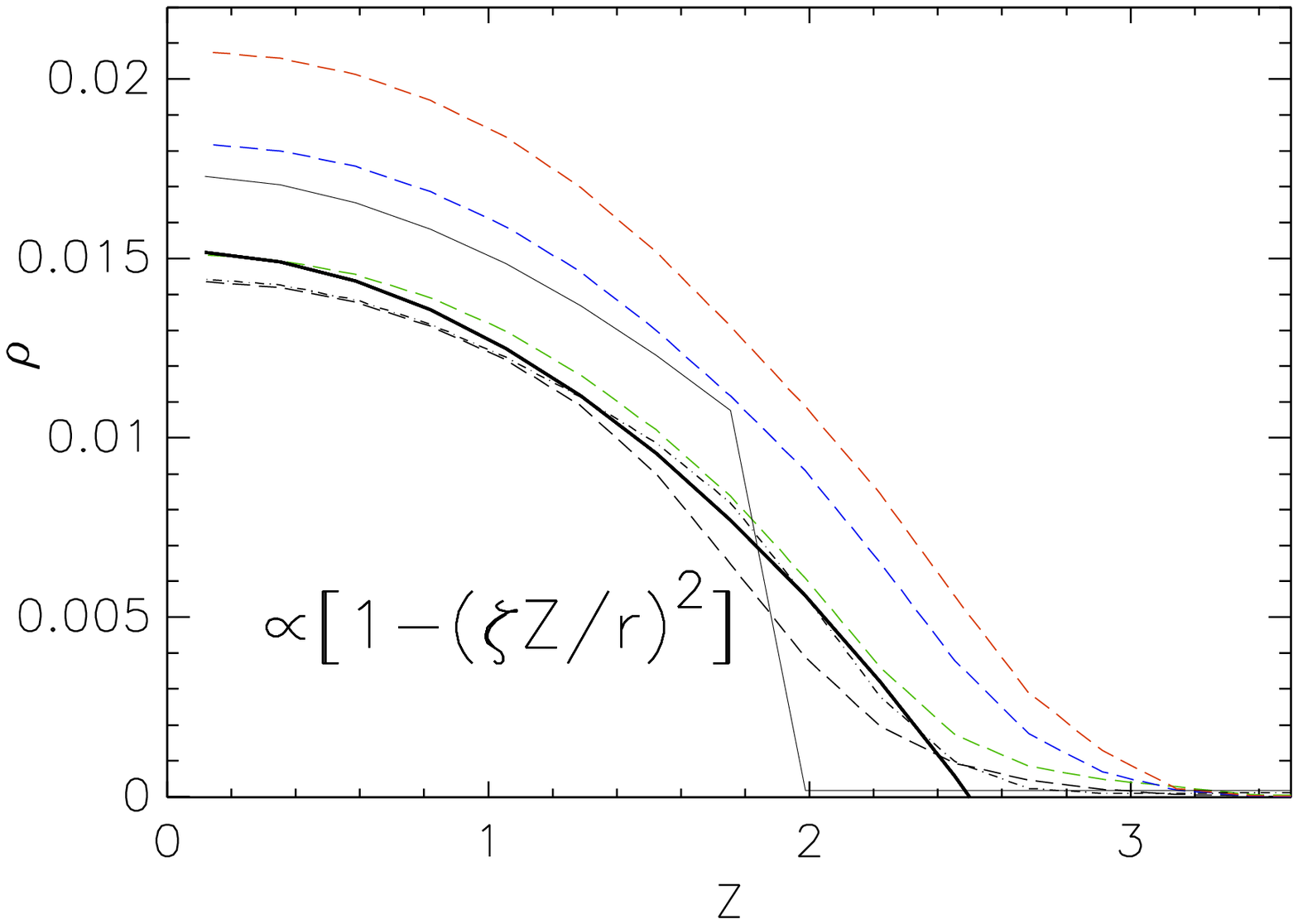}
\caption{Comparison of the matter density in the initial set-up (thin
solid line) with the quasi-stationary solutions in the numerical
simulations in the HD (dot-dashed line) and the MHD (long-dashed line)
cases, with $\Omega$=0.2$\Omega_{\rm br}$. Left panel: radial dependence
along the midplane, just above $\theta$=$90^\circ$. Right panel: the
profiles along the vertical line at r=15$R_\star$. The HD and MHD
profiles are nearly identical. In black, green, blue and red colors are
the results in the MHD cases with the stellar magnetic field strengths
0.25, 0.5, 0.75 and 1.0 kG, respectively (from bottom to top along the
line about the middle of the X-axis in both panels). The closest match to
the 0.5 kG case is depicted with the thick solid line.}
\label{fig:comparerho1}
\end{figure*}

\begin{figure*}
\centering
\includegraphics[width=\columnwidth]{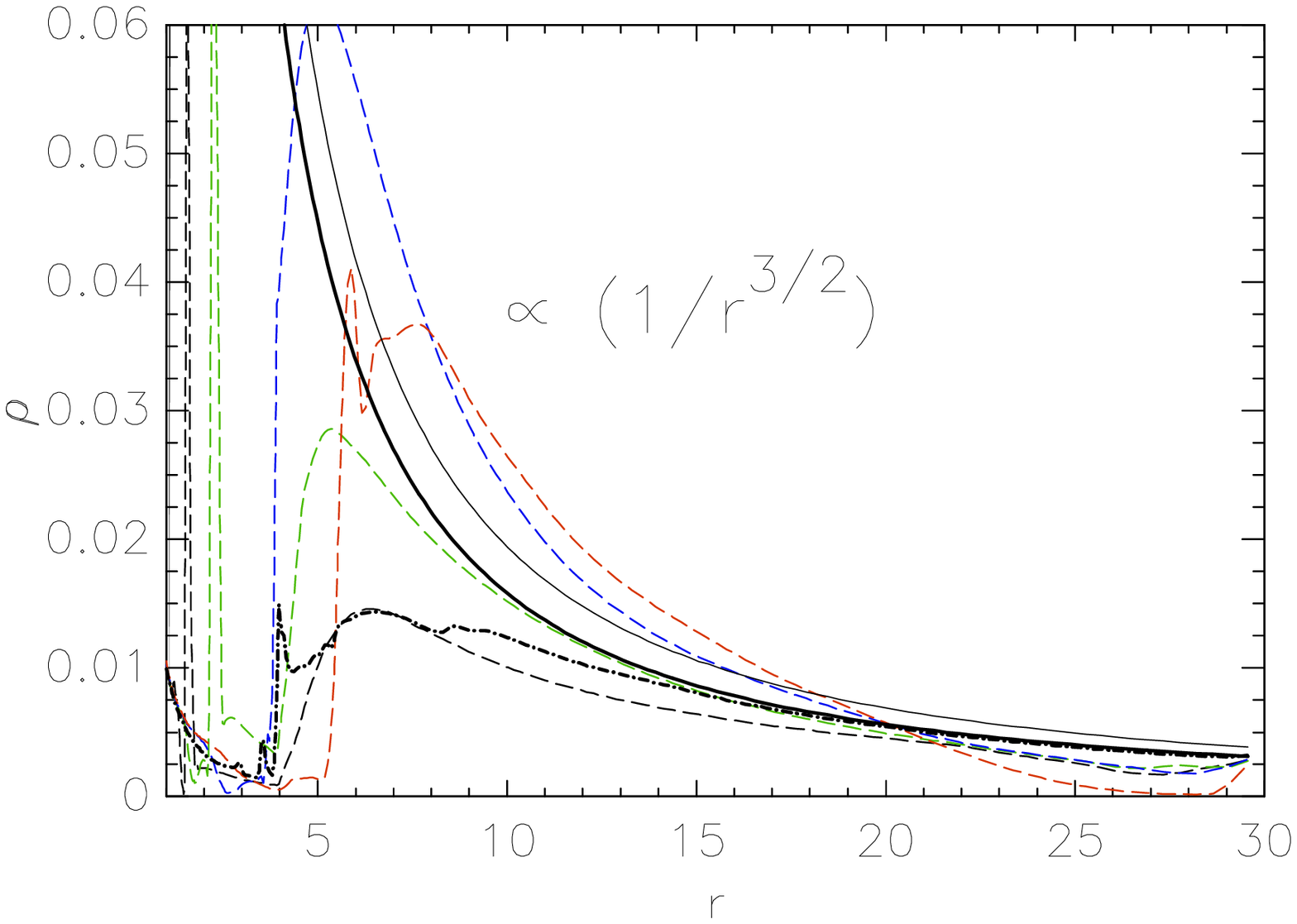}
\includegraphics[width=\columnwidth]{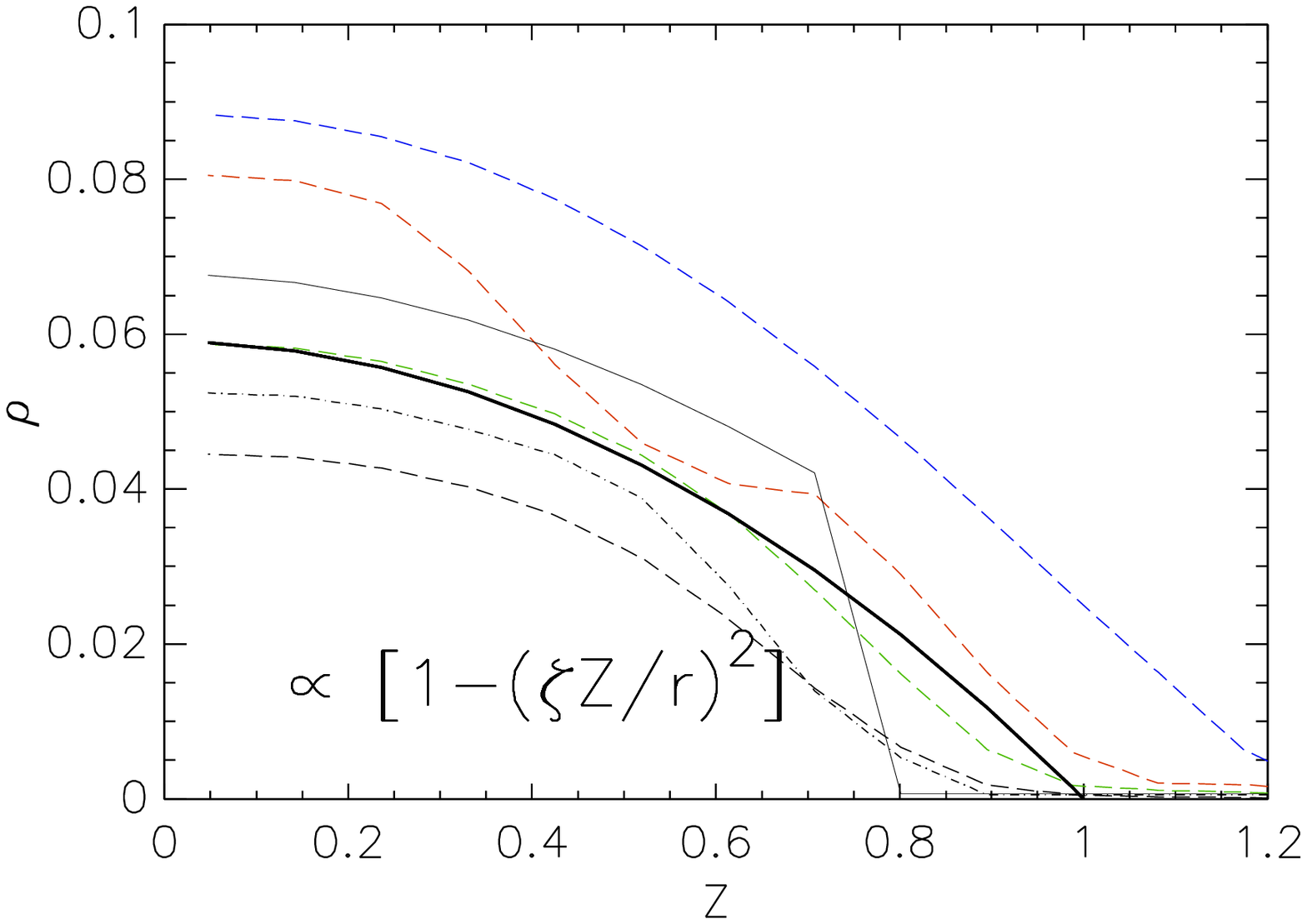}
\caption{Comparison of the matter density in the initial set-up (solid
line) with the quasi-stationary solutions in numerical simulations
in the HD (dot-dashed line) and the MHD cases with
$\Omega$=0.2$\Omega_{\rm br}$ along the disk surface (left panel)
at $\theta$=$83^\circ$, and along a vertical line at $r$=$6R_\star$
(right panel). The meaning of the lines is the same as in
Fig.~\ref{fig:comparerho1}.}
\label{rhocompareb}
\end{figure*}
%
We check now if the numerical solutions in the inner and outer disk
regions are compatible with the conditions derived from the analytical
equations. For the comparison, we use the expressions listed in the
Eqs.~\ref{compeqsa}-\ref{etasoba}.

Results for the radial dependence along a line just above the equatorial
mid-plane of the disk, and for the vertical dependence along a line at
$r$=15$R_\star$ are shown in Figs.~\ref{fig:comparerho1}, \ref{rhocompareb} and
Figs.~\ref{fig:compare1}-\ref{fig:compareetas} in Appendix. The
matching function for each physical quantity is also depicted.

To show that the matching function is of the same shape along the disk
surface at $\theta=83^\circ$ as it was along the equatorial line, in
the left panel in Fig.~\ref{rhocompareb} is shown the result for the
matter density, a similar result is obtained for the other physical
quantities.

An example of the matching function along a vertical direction closer to
the star than $r$=$15R_\star$ is shown in the right panel in
Fig.~\ref{rhocompareb}, with the density along a vertical direction at
$r$=$6R_\star$. Again, the matching function is of the same shape as
along a line further from the star, only the proportionality constant is
different. For the other physical quantities we obtain a similar result.

How do the obtained expressions compare to the general conditions in
\S~\ref{asec1} obtained from the analytical equations?

$\bullet$ The numerical solution for the density in the magnetic case has
the same dependence as the analytical one in the HD case. Both can be
approximated by the same expression, with the difference only in the
proportionality constant.

$\bullet$ The same is true for the velocity components, with the difference
between the two numerical solutions most visible in the radial dependence in
radial and vertical components of the poloidal velocity. The azimuthal
velocity component does not change from the initial value since it is
not evolved in our two-dimensional axisymmetric simulations.

$\bullet$ The magnetic field components in the disk in the simulations
follow the expected $1/r^3$ decrease in the dipole field strength with
distance from the star.

$\bullet$ In the analytical solution, all three magnetic field components
are functions of $r$ alone in the zeroth and first order in $\epsilon$.
With the
nonvanishing magnetic field in the disk, and its vertical dependence on
$z$, this would lead to the conclusion that $\mathbf{B}_0=\mathbf{B}_1$=0,
and vertical, linear dependence on height above the disk equatorial
plane should be related to the higher, second order in $\epsilon$. Such
a result satisfies the ${\rm div}\mathbf{B}=0$ condition in the second order
in $\epsilon$, where we obtain a linear dependence in $B_{z2}= zf(r)$.
In our simulations all the components are proportional
to $z/r^3$, hinting to a similar analytical solution for $B_{r2}$ and
$B_{\varphi 2}$, with a linear dependence on $z$.

\section{Numerical solutions with different parameters}
What are the changes in our numerical solutions with the variation of the
physical parameters like the stellar rotation, magnetic field strength
and rotation rate, or the dissipation (viscous or resistive) in the disk?

Solutions with smaller stellar rotation rates follow similar trends, as
shown in \cite{cem19}.

Solutions with different strengths of the magnetic field are shown in
different colors in Figs.~\ref{fig:comparerho1} and
\ref{fig:compare1}-\ref{fig:compareetas} in
Appendix. The matching functions differ only in the proportionality
coefficients so that Eqs.~\ref{compeqsa}-\ref{compeqsb} are valid in the
cases with different fields. The coefficients in the cases with 500~G and
1000~G are listed in Table~\ref{tabsols}.

In the case of a weaker stellar field, 250~G, the geometry of the solution
is the same as in the presented case with 500~G, and in the case with
750~G. If the field in our simulation is increased to 1000~G, the magnetic
pressure pushes the disk inner rim away, and the accretion column is
unstable or even disrupted. The results in the parts of the disk which we
consider here are not affected by the change in the geometry at the inner
disk rim---the shape of the functions is still the same, only the
proportionality coefficients differ, as shown in the table above.

\section{Conclusions}
We use numerical simulations in combination with analytical conditions
obtained from the asymptotic approximation, to provide relations
describing a thin magnetic accretion disk.

Like in the HD case \citep{KK00}, we perform a Taylor expansion of the
equations of motion in the small parameter $\epsilon$, the disk thickness
to the radial dimension ratio. In addition to the equations of motion and
induction equation, for the first time we add the energy equation
into the asymptotic approximation. We obtain the zeroth, the first and
the second order terms in $\epsilon$. 

In the magnetic case, equations in the disk cannot be solved without
knowing the solution in the corona between the disk and the stellar
surface. It is because of the connectivity of the magnetic field
in the disk to the stellar surface. The solution is further
complicated with the magnetic reconnection taking place in the
corona. In effect, we can only derive a set of general conditions
which should be satisfied for a self-consistent solution of the
equations.

From star-disk magnetospheric interaction simulations we obtain
the quasi-stationary solutions for a magnetic geometrically thin disk.
We write a set of expressions representing the physical variables in the
disk. Such expressions are then compared with the general conditions
extracted from the asymptotic approximation solution. 

$\bullet$ We find  the velocity field in the MHD disk to be broadly
similar to the HD (numerical and analytical solutions),
with the vertical ($z$) component of velocity enhanced for low values of
the stellar magnetic field, and the radial ($r$) component enhanced for
high values of $B$, while the azimuthal ($\varphi$) component is insensitive
to the stellar magnetic field.

$\bullet$ The density in the disk has the same functional form in the HD
and MHD cases, with its value somewhat lower for high values of the
stellar magnetic field.

$\bullet$ Numerical solution for the density and velocity components
in the magnetic case follows the same dependence as the analytical
solution in the HD case. The difference is only in the proportionality
coefficients.

$\bullet$ Magnetic field components in the disk follow the $z/r^3$
dependence, where $r^{-3}$ is the expected radial decrease for the
stellar dipole field. The difference in the solutions with the
different magnetic field strengths is only in the proportionality
coefficients.

$\bullet$ We find that the results from numerical simulations satisfy
the conditions obtained from analytical equations. The expressions
matching the numerical solutions in the middle part of the disk are
valid in the cases with different stellar magnetic field strengths.
Only the coefficients of proportionality change.

$\bullet$ We compared here the analytical solutions with numerical
solutions in the cases with stellar rotation equal to 20\% of the
equatorial mass-shedding (``breakup'') velocity.
As shown in \cite{cem19}, solutions with smaller
stellar rotation rates follow similar trends, so that our conclusions
extend to such cases. 

Our study here is limited to the values of the free parameters of
viscosity and resistivity $\alpha_{\rm v}=\alpha_{\rm m}=1$. We leave
investigation of the solutions in other cases for a separate study,
in particular the case with a smaller viscosity parameter
$\alpha_{\rm v}<0.685$, which shows a backflow region in the disk close
to the disk equatorial plane. We also leave for a separate study the
cases with faster rotating stars, as
they often exhibit axial jets and conical outflows, changing the
geometry of the solutions. 

We performed simulations in a quadrant of the meridional plane, enforcing
the equatorial disk plane as a boundary condition. It remains to check
the difference from solutions in the full meridional plane.

\section*{Acknowledgements}
M\v{C} developed the setup for star-disk simulations while in CEA, Saclay,
under the ANR Toupies grant, and his collaboration with Croatian STARDUST
project through HRZZ grant IP-2014-09-8656 is also acknowledged. Work at
CAMK is funded by the Polish NCN grant 2013/08/A/ST9/00795, and VP
work is partly funded by the Polish National Science Centre grant
2015/18/E/ST9/00580. We thank IDRIS (Turing cluster) in Orsay, France,
ASIAA/TIARA (PL and XL clusters) in Taipei, Taiwan and NCAC (PSK and CHUCK
clusters) in Warsaw, Poland, for access to Linux computer clusters used for the
high-performance computations. The {\sc pluto} team is thanked for the
possibility to use the code. We thank CAMK Ph.D. student D. A. Bollimpalli
and summer students F. Bartoli\'{c} and C. Turski for developing the
Python scripts for visualization.


\appendix{
\section{Asymptotic approximation equations for a thin accretion disk}
We illustrate the asymptotic approximation method in detail by deriving
all the terms through the second order in the continuity equation.
The remaining equations are derived by following the same method.
We present second order equations of the set from Section 2. Unlike in
the HD case (KK00), in general these cannot be solved without additional
assumptions and/or boundary conditions.

From the reflection symmetry about the $z$=0 midplane of the disk it
follows that $\rho$, $P$, $c_s$, $\varv_r$ and $\varv_\varphi$, i.e. $\Omega$ are
even functions of $z$ under reflections through the equatorial plane, and
$\varv_z$ is the odd function of $z$ - see KK00. It is assumed that all the
terms in the expansion of any quantity are of the same parity. We follow
the same assumptions about the reflection symmetry for the hydrodynamic
quantities in the magnetic case, but we do not extend the assumption to
the magnetic field components.

In the hydrodynamic solution one can assume that the disk density
decreases smoothly to zero towards the disk surface, which greatly
simplifies the solution. In the magnetic case, the disk solution cannot
be given without inclusion of the stellar corona, because of a magnetic
connection with the star and a corona above the disk. To obtain a
solution for the magnetic field penetrating the disk, we have to include
the disk-corona boundary condition, which is unknown. Because of this, we
can obtain only the most general conditions for the disk magnetic field from
the equations. The information about the magnetic field solution inside
the disk we obtain from our numerical simulations.

We will be searching for the stationary solutions, so that the additional
conditions are that of stationarity, $\partial/\partial t=0$,
and the axial symmetry $\partial/\partial\varphi=0$. We work in the
cylindrical coordinates $(r, z, \varphi)$. The normalization is defined with
the following equations:
$\epsilon=\tilde{c}_{\mathrm s}/(\tilde{R}\tilde{\Omega})=\tilde{H}/\tilde{R}\ll 1$,
so that $\tilde{c}_{\mathrm s}=\epsilon\tilde{R}\tilde{\Omega}$,
and then
$c_s'=c_s/\tilde{c}_{\mathrm s}=c_s/(\epsilon\tilde{R}\tilde{\Omega})$.
Twiddles denote characteristic values of the variables,
and primes the scaled variables. Further,
$\Omega'=\Omega/\tilde{\Omega}$,
$\tilde{\Omega}=\Omega_{\mathrm K}=\sqrt{GM_\star/\tilde{R}^3}$,
$r'=r/\tilde{R}$, $z'=z/(\epsilon\tilde{R})$,
$\varv_r'=\varv_r/\tilde{c}_{\mathrm s}=\varv_r/(\epsilon\tilde{R}\tilde{\Omega})$,
$\varv_z'=\varv_z/\tilde{c}_{\mathrm s}=\varv_z/(\epsilon\tilde{R}\tilde{\Omega})$,
$\varv_\varphi'=\varv_\varphi/\tilde{c}_{\mathrm s}=\varv_\varphi/(\tilde{R}\tilde{\Omega})$.
The magnetic field we normalize with the Alfv\'{e}n speed
$\tilde\varv_{\mathrm A}=\tilde{B}/\sqrt{4\pi\tilde{\rho}}$ as a characteristic
speed, and $\rho'=\rho/\tilde{\rho}$. Then we have
$B'=B/\tilde{B}=B/({\tilde\varv}_{\mathrm A}^2\sqrt{4\pi\tilde{\rho}})$, and
$\tilde{B}$ is the normalization for all the magnetic field components:
$B_r'=B_r/\tilde{B}$, $B_z'=B_z/\tilde{B}$,
$B_\varphi'=B_\varphi/\tilde{B}$.

The beta plasma parameter
$\beta=P_{\mathrm {gas}}/P_{\mathrm {mag}}=8\pi P_{\mathrm {gas}}/B^2$. With
$P=P_{\mathrm {gas}}$ we can write
$c_s^2=\gamma P/\rho=\gamma P B^2/(8\pi\rho)=\gamma\beta B\varv_{\mathrm A}/2$,
so that $\tilde{\varv}_{\mathrm A}^2/\tilde{c}_{\mathrm s}^2=2/(\gamma\tilde{\beta})$.

The viscosity scales with the sound speed as a characteristic velocity and
the height of the disk $H$, so that the normalization for the kinetic
viscosity is $\tilde{\nu}_{\mathrm
v}=\tilde{c}_{\mathrm s}\tilde{H}=\epsilon^2\tilde{R}^2\tilde{\Omega}$,
and then $\tilde{\eta}=\tilde{\rho}\tilde{\nu}_{\mathrm v}=
\tilde{\rho}\epsilon^2\tilde{R}^2\tilde{\Omega}$. Then
$\eta'=
\eta/\tilde{\eta}=\eta/(\tilde{\rho}\epsilon^2\tilde{R}^2\tilde{\Omega})$.
For the resistivity we choose the normalization with the Alfv\'{e}n speed as a
characteristic speed, so that
$\tilde{\eta}_{\mathrm m}=\tilde{\varv}_{\mathrm A}\tilde{H}=
\epsilon\tilde{R}\tilde{\varv}_{\mathrm A}$.
Then $\eta'_{\mathrm m}=\eta_{\mathrm m}/\tilde{\eta}_{\mathrm m}=
\eta_{\mathrm m}/(\epsilon^2\tilde{R}\tilde{\varv}_{\mathrm A})=
\eta_{\mathrm m}\left(
\sqrt{\gamma\tilde{\beta}/2}\right)/(\tilde{c}_{\mathrm s}\epsilon\tilde{R})
=\eta_{\mathrm m}\left(
\sqrt{\gamma\tilde{\beta}/2}\right)/(\epsilon^2\tilde{R}^2\tilde{\Omega})$.

In the asymptotic approximation, we write all the variables in the Taylor
expansion with the coefficient of expansion $\epsilon=\tilde{H}/\tilde{R}<<1$
(see KK00). For a variable X, we then have
$X=X_0+\epsilon X_1+\epsilon^2 X_2+\epsilon^3 X_3+\dots $, and we can compare
the terms of the same order in $\epsilon$. Omitting primes in the normalized
variables, we write the normalized equations of continuity, magnetic field
solenoidality ($\nabla\cdot\mathbf{B}=0$), momentum, induction and energy
density. For simplicity, in some cases we use the notation
$\partial_x=\partial/\partial x$, and we drop all primes in the following
(where all the variables are scaled, so no confusion can arise). 

\subsection*{Equation of continuity}
We start from the continuity equation:
\beq
\frac{\partial\rho}{\partial t}+\nabla\cdot(\rho\mathbf{v})=0.
\eeq
In the stationary case, when $\partial_t\rho=0$, and applying also the
axi-symmetry condition $\partial_\varphi(\rho\mathbf{v})=0$:
\begin{equation}
  \frac{1}{r}\partial_r(r\rho\varv_r)+
\partial_z(\rho\varv_z)=0.\nonumber
\end{equation}
We can write the
normalized equation, in which the terms can be written in the orders of a
small parameter $\epsilon$:
\begin{equation}
\frac{1}{\tilde{R}r'}\frac{1}{\tilde{R}}\partial_{r'}
(r'\tilde{R}\tilde{\rho}\rho'\epsilon\tilde{\Omega}\tilde{R}\varv_{r'})+
\frac{1}{\epsilon\tilde{R}}\partial_{z'}\rho\tilde{\rho}\epsilon\tilde{\Omega}\tilde{R}\varv_{z'}
=0\ \big/ \frac{1}{\tilde{R}\tilde{\rho}}.\nonumber
\end{equation}
Removing the primes, we can write:
\begin{equation}
\frac{\epsilon}{r}\partial_r(r\rho\varv_r)+\partial_z(\rho\varv_z)=0.\nonumber
\end{equation}
Writing the expansion in $\epsilon$ in each quantity, we obtain
\begin{eqnarray}
\frac{\epsilon}{r}\partial_r[r(\rho_0+\epsilon\rho_1+\epsilon^2\rho_2+\dots)(\varv_{r0}+\epsilon\varv_{r1}+\epsilon^2\rho_2+\dots)]\nonumber \\
+\partial_z[(\rho_0+\epsilon\rho_1+\epsilon^2\rho_2+\dots)(\varv_{z0}+\epsilon\varv_{z1}+\epsilon^2\varv_{z2}+\dots)]=0.\nonumber
\end{eqnarray}
From this we can write the term in the order zeroth order in $\epsilon$ as:
\subsubsection*{Order $\epsilon^{0}$:}
\begin{equation}
\frac{\partial}{\partial z}\left(\rho_0\varv_{z0}\right) = 0 \ \Rightarrow\varv_{z0} = 0.\nonumber
\end{equation}
Since $\rho_0\ne 0$ is an even function, and $\varv_z$ is odd with respect to z, at the disk
equatorial plane this product is $\rho_0\varv_{z0}=0$. Since it does not depend on z,
we conclude that it must be $\varv_{z0}=0$.
\subsubsection*{Order $\epsilon^{1}$:}
In the first order in $\epsilon$ we have:
\begin{equation}
\frac{1}{r}\frac{\partial}{\partial r}\left(r\rho_0\varv_{r0}\right) + \frac{\partial}
{\partial z}\left(\rho_0\varv_{z1}\right) = 0\ \Rightarrow\varv_{z1}=0.\nonumber
\end{equation}
As we will see from the first order in
$\epsilon$ of the radial momentum, eq.~(\ref{radmom1}), we have $\varv_{r0}=0$, so that here we have
$\partial_z(\rho_0\varv_{z1})=0\Rightarrow \rho_0\varv_{z1}={\rm const}$ along $z$. Since
$\varv_z$ is odd with respect to z, following the same argumentation as above,
we conclude that $\varv_{z1}=0$.

\subsubsection*{Order $\epsilon^{2}$:}
In the second order in $\epsilon$ we have:
\begin{equation}
\frac{1}{r}\frac{\partial}{\partial r}\left(r\rho_0\varv_{r1}\right)+
\frac{\partial}{\partial z}\left(\rho_0\varv_{z2}\right) = 0.\nonumber
\end{equation}
The same procedure is carried in each of the following equations.

In the following, we will often find that  certain quantities are functions of the radial variable alone. In such cases we will denote a {\it generic} radial function as $f(r)$, without implying any particular functional dependence on $r$, so that the results $a =f(r)$, and $b =f(r)$ do {\it not} imply $a(r,z)\equiv b(r,z)$.

\subsection*{Condition ${\bf div}\mathbf{B}{\bf =0}$:}
\begin{equation}
\frac{\epsilon}{r}\frac{\partial}{\partial r}\left(r B_r\right) + \frac{\partial}{\partial z}
\left(B_z\right) = 0
\end{equation}

\subsubsection*{Order $\epsilon^{0}$:}
\begin{equation}
\frac{\partial B_{z0}}{\partial z} = 0 \ \Rightarrow B_{z0}=f(r)
\label{bzofr}
\end{equation}

\subsubsection*{Order $\epsilon^{1}$:}
\begin{equation}
  \frac{1}{r}\frac{\partial}{\partial r}\left(r B_{r0}\right)
  + \frac{\partial}{\partial z}\left(B_{z1}\right) = 0\Rightarrow B_{z1}=f(r)
\label{divb2}
\end{equation}
From the first order in $\epsilon$ in the azimuthal component of the
induction equation we have that $B_{r0}$=0, so that
$B_{z1}=f(r)$.

\subsubsection*{Order $\epsilon^{2}$:}
\begin{equation}
\frac{1}{r}\frac{\partial}{\partial r}\left(r B_{r1}\right) + \frac{\partial}{\partial z}
\left(B_{z2}\right) = 0
\label{divb3}
\end{equation}
From the later equations we will show that $B_0=f(r)$ and $B_1=f(r)$ or 0,
together with all their components, so that we can integrate the above
equation in $z$, to obtain
\beq
B_{z2}\Big|^{z}_{0}=-\frac{\partial_r(rB_{r1})}{r}z\Big|^{z}_{0}= zf(r),
\label{linb2}
\eeq
with a linear dependence in the vertical direction.

\subsection*{Radial momentum:}
\begin{equation}
\begin{aligned}
\epsilon^2\varv_r\frac{\partial\varv_r}{\partial r}+\epsilon\varv_z\frac{\partial\varv_r}{\partial z}-\Omega^2r =-\frac{1}{r^2}
\left[1+\epsilon^2\left(\frac{z}{r}\right)^2\right]^{-3/2}\\
-\epsilon^2n\frac{\partial c_{\mathrm s}^2}{\partial r}+\frac{2}{\gamma\tilde{\beta}}\frac{1}{\rho}\left(\epsilon^2B_r\frac{\partial B_r}{\partial r}
+\epsilon B_z\frac{\partial B_r}{\partial z}-\epsilon^2\frac{B_\varphi^2}{r}\right)\\
-\frac{\epsilon^2}{\gamma\tilde{\beta}}\frac{1}{\rho}\frac{\partial B^2}{\partial r}+ 
\frac{\epsilon^3}{\rho r}\frac{\partial}{\partial r}\left(2\eta r\frac{\partial\varv_r}{\partial r}\right)+\frac{\epsilon}{\rho}\frac{\partial}{\partial z}\left(\eta\frac{\partial\varv_r}{\partial z}\right)\\
+\frac{\epsilon^2}{\rho}\frac{\partial}{\partial z}\left(\eta\frac{\partial\varv_z}{\partial r}\right)-\epsilon^3\frac{2\eta\varv_r}{\rho r^2}-\frac{2\epsilon^3}{3\rho}\frac{\partial}{\partial r}\left(\eta\frac{1}{r}\frac{\partial}{\partial r}\left(r\varv_r\right)\right)\\
-\frac{2}{3}\frac{\epsilon^2}{\rho}\frac{\partial}{\partial r}\left(\eta\frac{\partial\varv_z}{\partial z}\right).
\end{aligned}
\end{equation}
Here $n$ is the polytropic index. In the case of adiabatic index $\gamma=5/3$
for an ideal gas, we have $n=3/2$.
\subsubsection*{Order $\epsilon^{0}$:}
\begin{equation}
\Omega_{0} = r^{-3/2}
\label{omega0}
\end{equation}

\subsubsection*{Order $\epsilon^{1}$:}
\begin{equation}
\begin{aligned}
  - 2r\Omega_{0}\Omega_{1} =
  \frac{2}{\gamma \tilde{\beta}}\frac{1}{\rho_{0}} B_{z0}
  \frac{\partial B_{r0}}{\partial z} +
  \frac{1}{\rho_{0}} \frac{\partial}{\partial z}
  \left(\eta_0 \frac{\partial \varv_{r0}}{\partial z}\right)\\
   \Rightarrow \frac{\partial B_{r0}}{\partial z}=0.
\label{radmom1}
\end{aligned}
 \end{equation}

From the vertical symmetry $\Omega_1=0$ as shown in KK00, see also Appendix
A in \citet{Reb09} for more formal derivation. In the HD case, then,
$\varv_{r0}=0$ and if this is maintained in the MHD case, we have
$\partial_z B_{r0}=0 \Rightarrow B_{r0}=f(r)$.

\subsubsection*{Order $\epsilon^{2}$:}
\begin{equation}
\begin{aligned}
2r\rho_0\Omega_0\Omega_2=\frac{3\rho_0}{2}\frac{z^2}{r^4}+n\rho_0\frac{\partial c_{s0}^2}{\partial r}-\frac{\partial}{\partial z}\left(\eta_0\frac{\partial\varv_{r1}}{\partial z}\right)\\
-\frac{2}{\gamma\tilde{\beta}}\left( B_{r0}\frac{\partial B_{r0}}{\partial r}+B_{z0}\frac{\partial B_{r1}}{\partial z}-\frac{B_{\varphi 0}^2}{r}\right)+\frac{1}{\gamma\tilde{\beta}}\frac{\partial B_0^2}{\partial r}
\label{radmom2}
\end{aligned}
\end{equation}
  
\subsection*{Azimuthal momentum:}
\begin{equation}
\begin{aligned}
\epsilon \frac{\rho\varv_r}{r^2}\frac{\partial}{\partial r}\left(r^{2}\Omega\right) + 
\rho \varv_z \frac{\partial \Omega}{\partial z} = \frac{\epsilon^2}{r^3}\frac{\partial}{\partial r}
\left(r^3\eta\frac{\partial \Omega}{\partial r}\right) \\
+\frac{\partial}{\partial z}\left(\eta \frac{\partial \Omega}{\partial z}\right)+\frac{2}{\gamma\tilde{\beta}}
\frac{1}{r}\left(\epsilon^2 B_r\frac{\partial B_{\varphi}}{\partial r} + \epsilon B_z\frac{\partial B_\varphi}{\partial z} +
\epsilon^2\frac{B_\varphi B_r}{r}\right)
\end{aligned}
\end{equation}

\subsubsection*{Order $\epsilon^{0}$:}
\begin{equation}
0 = \frac{\partial}{\partial z}\left(\eta_0\frac{\partial \Omega_{0}}{\partial z}\right),
\label{azmom}
\end{equation}
consistent with Eq.~(\ref{omega0}).

\subsubsection*{Order $\epsilon^{1}$:}
\begin{equation}
\begin{aligned}
\frac{\rho_{0}\varv_{r0}}{r^{2}}\frac{\partial}{\partial r}\left(r^{2}\Omega_0\right) = 
\frac{\partial}{\partial z}\left(\eta_0\frac{\partial \Omega_1}{\partial z}\right) + 
\frac{2}{\gamma \tilde{\beta}}\frac{1}{r}B_{z0}\frac{\partial B_{\varphi 0}}{\partial z}\\
\Rightarrow \frac{\partial B_{\varphi 0}}{\partial z}=0.
\end{aligned}
\end{equation}
Since $\varv_{r0}=0$ and $\Omega_1=0\ \Rightarrow\partial_zB_{\varphi_{0}}=0$.
This matches the conclusion from the zeroth order in $\epsilon$ in the
energy equation below, that $B_{\varphi_{0}}=f(r)$.

\subsubsection*{Order $\epsilon^{2}$:}
\begin{equation}
\begin{aligned}
\frac{\rho_{0}\varv_{r1}}{r}\frac{\partial}{\partial r}\left(r^{2}\Omega_0\right)= 
\frac{2}{\gamma \tilde{\beta}}\left(B_{r0}\frac{\partial B_{\varphi 0}}
{\partial r}+B_{z0}\frac{\partial B_{\varphi 1}}{\partial z}+\frac{B_{r0}B_{\varphi
0}}{r}\right)\\
+\frac{1}{r^2}\frac{\partial}{\partial r}
\left(r^3\eta_0\frac{\partial \Omega_0}{\partial r}\right)
+\frac{\partial}{\partial z}\left(\eta_0\frac{\partial\Omega_2}{\partial z}\right) .
\label{azmom2}
\end{aligned}
\end{equation}

\subsection*{Vertical momentum:}
\begin{equation}
\begin{aligned}
\epsilon \varv_r \frac{\partial \varv_z}{\partial r} + \varv_z \frac{\partial
\varv_z}{\partial z} =-\frac{z}{r^{3}}\left[1+\epsilon^{2}\left(\frac{z}{r}\right)^{2}\right]^{-3/2}\\
-n\frac{\partial c_{\mathrm s}^2}{\partial z}+\frac{2}{\gamma\tilde{\beta}}\frac{1}{\rho}\left(\epsilon B_r\frac{\partial B_z}
{\partial r}+B_z\frac{\partial B_z}{\partial z}\right)-\frac{1}{\gamma\tilde{\beta}}
\frac{1}{\rho}\frac{\partial B^{2}}{\partial z}\\
+\frac{2}{\rho}\frac{\partial}{\partial z}\left(\eta \frac{\partial \varv_z}{\partial z}\right)
+\frac{\epsilon^{2}}{\rho r}\frac{\partial}{\partial r}\left(r \eta \frac{\partial \varv_z}{\partial r}\right)\\
-\frac{2}{3}\frac{\epsilon}{\rho}\frac{\partial}{\partial z}\left(\frac{\eta}{r}\frac{\partial}{\partial r}\left(r\varv_r\right) \right)-\frac{2}{3\rho}\frac{\partial}{\partial z}\left(\eta \frac{\partial \varv_z}{\partial z}\right)\\
+\frac{\epsilon}{\rho r} \frac{\partial}{\partial r}\left(\eta r\frac{\partial \varv_r}{\partial z}\right)
\end{aligned}
\end{equation}

\subsubsection*{Order $\epsilon^{0}$:}
\begin{equation}
    0 = -\frac{z}{r^{3}} - n \frac{\partial c_{\rm s{0}}^{2}}{\partial z} - \frac{1}{\gamma 
\tilde{\beta}}\frac{1}{\rho_{0}}\frac{\partial B_{0}^{2}}{\partial z} 
\end{equation}
Since we had $\partial B_{r0}/\partial z=\partial B_{z0}/\partial z=\partial B_{\varphi 0}/\partial z=0$, we have
$\partial B_0/\partial z=0$, i.e. $B_0=f(r)$. We have then
\begin{equation}
\frac{z}{r^{3}}= - n \frac{\partial c_{\rm s{0}}^{2}}{\partial z},
\label{vertmom0}
\end{equation}
which is the vertical hydrostatic equilibrium equation.

The disk solution in \citet{H77} and KK00 has been obtained by assuming that
the disk density decreases towards the surface, $\rho_0\rightarrow0$.
If, instead, we supply at the disk
surface a value at the boundary with the coronal density $\rho_{\rm cd}$,
we obtain:
\beq
\rho_0=\left[\rho_{\rm cd}^{2/3}+\frac{h^2-z^2}{5r^3}\right]^{3/2},
\label{rhozero}
\eeq 
where $h$ is the disk semi-thickness.
The pressure and sound speed now become:
\beq
P_0=\left[\rho_{\rm cd}^{2/3}+\frac{h^2-z^2}{5r^3}\right]^{5/2}, c_{\rm s{0}}=
\sqrt{\frac{5}{3}\left[\rho_{\rm cd}^{2/3}+\frac{h^2-z^2}{5r^3}\right]}.
\label{pressound}
\eeq
The \citet{H77} solution is recovered by setting $\rho_{\rm cd}=0$, for the
boundary at the disk maximal height.

In our case, since $h\propto r$, we
can write, with the proportionality constant $h'$,  $h=h'r$. Assuming
the corona at the surface of the disk to be in the hydrostatic equilibrium,
with $\rho_{\mathrm cd}\propto(\rho_{\mathrm c0}/r)^{3/2}$ we can write:
\begin{equation}
\begin{aligned}
c_{\mathrm s0}^2=\frac{5}{3}
\left(\rho_{\mathrm cd}^{2/3}+\frac{h'^2r^2-z^2}{5r^3}\right)=\frac{5}{3}
\left(\frac{k_\rho\rho_{\mathrm c0}}{r}+\frac{h'^2}{5r}-\frac{z^2}{5r^3}\right)\\
=\frac{5k_\rho\rho_{\mathrm c0}+h'^2}{3r}\left[1-\left(\zeta\frac{z}{r}\right)^2\right]
\propto\frac{1}{r}\left[1-\left(\zeta\frac{z}{r}\right)^2\right],
\end{aligned}
\label{cskvadr}
\end{equation}
with $\zeta^2=1/(5k_\rho\rho_{\mathrm c0}+h'^2)$, where
$k_\rho$ is the proportionality constant, and
$\rho_{\mathrm c0}\sim0.01$ is the ratio between the initial corona
and disk density.

\subsubsection*{Order $\epsilon^{1}$:}
\begin{equation}
\begin{aligned}
0 = - 2n\rho_0\frac{\partial}{\partial z}\left(c_{\mathrm s0}c_{\mathrm s1}\right)-n\rho_1\frac{\partial c_{\mathrm s0}^2}{\partial z}\\
+\frac{2}{\gamma\tilde{\beta}}\left(B_{r0}\frac{\partial B_{z0}}{\partial r}+B_{z0}\frac{\partial B_{z1}}{\partial z}\right)-\frac{1}{\gamma\tilde{\beta}}\frac{\partial}{\partial z}\left(2B_0B_1\right)\\
+\frac{4}{3}\frac{\partial}{\partial z}\left(\eta_0\frac{\partial\varv_{z1}}{\partial z}\right)-\frac{2}{3}\frac{\partial}{\partial
z}\left(\frac{\eta_0}{r}\frac{\partial}{\partial r}(r\varv_{r0})\right)
\end{aligned}
\label{vertmom1}
\end{equation}
Since $c_{\mathrm s1}$=$\rho_1$=$\varv_{z1}$=$\varv_{r0}$=0 , this leaves us with:
\begin{equation}
B_{r0}\frac{\partial B_{z0}}{\partial r}+B_{z0}\frac{\partial B_{z1}}{\partial z}=B_0\frac{\partial B_1}{\partial z}
\label{vertmom1b}
\end{equation}
With $B_{r0}=0$, which we obtain in the azimuthal component of the
induction equation, Eq.~(\ref{azind0}), and $B_{z1}=f(r)$ obtained from Eq.~(\ref{divb2}), we
stay with:
\beq
\frac{\partial B_1}{\partial z}=0\Rightarrow B_1=f(r).
\label{bzobz1}
\eeq
 
\subsubsection*{Order $\epsilon^{2}$:}
\begin{equation}
\begin{aligned}
\frac{3\rho_0}{2}\frac{z^3}{r^7}=2n\rho_0\frac{\partial}{\partial z}\left(c_{\mathrm s0}c_{\mathrm s2}\right)
+n\rho_2\frac{\partial c_{\mathrm s0}^2}{\partial z}\\
-\frac{2}{\gamma\tilde{\beta}}\left(B_{r1}\frac{\partial B_{r0}}{\partial r}+B_{r0}\frac{\partial B_{r1}}{\partial r}+B_{z0}\frac{\partial B_{z2}}{\partial z}+B_{z1}\frac{\partial B_{z1}}{\partial z}\right)\\
+\frac{1}{\gamma\tilde{\beta}}\left(\frac{\partial B_1^2}{\partial z}+2B_0\frac{\partial B_2}{\partial z}\right)\\
-\left[\frac{1}{r}\frac{\partial}{\partial r}\left(r\eta_0\frac{\partial\varv_{r1}}{\partial z}\right)+2\frac{\partial}{\partial z}\left(\eta_0\varv_{z2}\right)\right]\\
+\frac{2}{3}\frac{\partial}{\partial z}\left(\varv_{r1}+r\eta_0\frac{\partial\varv_{r1}}{\partial r}+\eta_0\frac{\partial\varv_{z2}}{\partial z}\right).
\end{aligned}
\label{vertmom2}
\end{equation}

\subsection*{Radial induction:}
\begin{equation}
\begin{aligned}
B_z\frac{\partial\varv_r}{\partial z}-\epsilon\varv_r \frac{\partial B_r}{\partial r}
 - \varv_z\frac{\partial B_r}{\partial z}-B_r\frac{\partial\varv_z}{\partial z}-\frac{\epsilon}{r}B_r\varv_r \\
 + \sqrt{\frac{2}{\gamma \tilde{\beta}}}\left(\frac{\partial \eta_{\rm m}}
{\partial z}\frac{\partial B_r}{\partial z}-\epsilon \frac{\partial \eta_{\rm m}}
{\partial z}\frac{\partial B_z}{\partial r}\right)\\
+\eta_{\rm m}\sqrt{\frac{2}{\gamma \tilde{\beta}}}\left(\frac{\epsilon^2}{r}\frac{\partial B_r}
{\partial r}+\epsilon^2\frac{\partial^2 B_r}{\partial r^2}-\frac{\epsilon^2}{r^2}B_r + 
\frac{\partial^2 B_r}{\partial z^2}\right)=0
\end{aligned}
\end{equation}

\subsubsection*{Order $\epsilon^{0}$:}
\begin{equation}
0 = B_{z0}\frac{\partial \varv_{r0}}{\partial z} + \sqrt{\frac{2}{\gamma \tilde{\beta}}}
\frac{\partial}{\partial z}\left(\eta_{\rm m0} \frac{\partial B_{r0}}{\partial z}\right)\Rightarrow 0=0.
\label{radind1}
\end{equation}

\subsubsection*{Order $\epsilon^{1}$:}
\begin{equation}
\begin{aligned}
0 = B_{z0}\frac{\partial \varv_{r1}}{\partial z}
+\sqrt{\frac{2}{\gamma\tilde{\beta}}}\frac{\partial}{\partial z}
\left(\eta_{\rm m0}\frac{\partial B_{r1}}{\partial z}\right)\\
-\sqrt{\frac{2}{\gamma\tilde{\beta}}}\frac{\partial \eta_{\rm m0}}{\partial z}
\frac{\partial B_{z0}}{\partial r}\Rightarrow 0=0.
\end{aligned}
\label{indRadialeps1}
\end{equation}
No new constraints.

\subsubsection*{Order $\epsilon^{2}$:}
\begin{equation}
\begin{aligned}
\varv_{r1}\frac{\partial B_{r0}}{\partial r}\sqrt{\frac{\gamma\tilde{\beta}}{2}}=\\
\frac{\partial\eta_{\rm m0}}{\partial z}\left(\frac{\partial B_{r2}}{\partial z}-\frac{\partial B_{z1}}{\partial r}\right)+\frac{\partial\eta_{\rm m1}}{\partial z}\left(\frac{\partial B_{r1}}{\partial z}-\frac{\partial B_{z0}}{\partial r}\right)\\
+\eta_{\rm m0}\left(\frac{1}{r}\frac{\partial B_{r0}}{\partial r}
-\frac{B_{r0}}{r^2}+\frac{\partial^2 B_{r0}}{\partial r^2}+\frac{\partial^2 B_{r2}}{\partial z^2}\right)\\
  +\eta_{\rm m1}\frac{\partial^2 B_{r1}}{\partial z^2}
\end{aligned}
\label{indRadialeps2}
\end{equation}

\subsection*{Azimuthal induction:}
\begin{equation}
\begin{aligned}
0 = \epsilon rB_r\frac{\partial\Omega}{\partial r}+rB_z\frac{\partial\Omega}{\partial z}-\epsilon^2\varv_r
\frac{\partial B_\varphi}{\partial r}-\epsilon\varv_z\frac{\partial B_\varphi}{\partial z}\\
-\epsilon^{2} B_{\varphi} \frac{\partial \varv_r}{\partial r}-\epsilon B_{\varphi} \frac{\partial \varv_z}{\partial z}\\
+\sqrt{\frac{2}{\gamma \tilde{\beta}}}\left(\frac{\epsilon^3}{r}\frac{\partial \eta_{\mathrm m}}{\partial r}\frac{\partial(rB_\varphi)}{\partial r}
+\epsilon\frac{\partial\eta_{\mathrm m}}{\partial z}\frac{\partial B_\varphi}{\partial z}\right)\\
+\sqrt{\frac{2}{\gamma \tilde{\beta}}}\eta_{\mathrm m}\left(\epsilon^3\frac{\eta_{\mathrm m}}{r}\frac{\partial}
{\partial r}B_\varphi+\epsilon^3\frac{\partial^2B_\varphi}{\partial r^2} +\epsilon\frac{\partial^2B_\varphi}{\partial z^2}-\epsilon^3 \frac{B_\varphi}{r^2}\right)
\end{aligned}
\label{azind}
\end{equation}

\subsubsection*{Order $\epsilon^{0}$:}
\begin{equation}
0 = rB_{z0}\frac{\partial \Omega_{0}}{\partial z}\Rightarrow 0=0.
\end{equation}

\subsubsection*{Order $\epsilon^{1}$:}
\beqa
0 = -\frac{3}{2}\Omega_{0}B_{r0} + rB_{z0}\frac{\partial \Omega_1}{\partial z}+\sqrt{\frac{2}{\gamma \tilde{\beta}}}
\frac{\partial}{\partial z}\left(\eta_{\rm m0}\frac{\partial B_{\varphi_0}}{\partial z}\right)
\label{azind0}
\eeqa
which gives $B_{r0}=0$.

\subsubsection*{Order $\epsilon^{2}$:}
\beqa
\begin{aligned}
\frac{3}{2}\frac{B_{r1}}{r^{3/2}}-rB_{z0}\frac{\partial\Omega_2}{\partial z}=\\
\sqrt{\frac{2}{\gamma\tilde{\beta}}}\left(\frac{\partial\eta_{\mathrm m0}}{\partial z}
\frac{\partial B_{\varphi 1}}{\partial z}+
\eta_{\mathrm m0}\frac{\partial^2 B_{\varphi 1}}{\partial z^2}\right).
\end{aligned}
\label{azind2}
\eeqa

\subsection*{Vertical induction:}
\begin{equation}
\begin{aligned}
0 = \epsilon B_r\frac{\partial \varv_z}{\partial r} - \epsilon \varv_r \frac{\partial B_z}
{\partial r} - \varv_z\frac{\partial B_z}{\partial z} -\epsilon\frac{B_z\varv_r}{r}\\
-\epsilon B_z\frac{\partial \varv_r}{\partial r} + \sqrt{\frac{2}{\gamma \tilde{\beta}}}
\left(\epsilon^{2}\frac{\partial \eta_{\mathrm m}}{\partial r}\frac{\partial B_z}{\partial r}
-\epsilon \frac{\partial \eta_{\mathrm m}}{\partial r}\frac{\partial B_r}{\partial z}\right) \\
+\sqrt{\frac{2}{\gamma \tilde{\beta}}}\eta_{\mathrm m}
\left(\epsilon^2\frac{1}{r}\frac{\partial B_z}{\partial r}
+\epsilon^{2}\frac{\partial^2B_z}{\partial r^{2}}+\frac{\partial^{2}B_z}{\partial z^{2}}\right)
\end{aligned}
\end{equation}

\subsubsection*{Order $\epsilon^{0}$:}
\begin{equation}
0 =  \sqrt{\frac{2}{\gamma \tilde{\beta}}} \eta_{\rm m0} 
\frac{\partial^{2}B_{z0}}{\partial z^{2}}\Rightarrow 0=0,
\end{equation}
which is in agreement with the previously obtained
$\partial B_{z0}/\partial z=0$.
 
\subsubsection*{Order $\epsilon^{1}$:}
\begin{equation}
0=\sqrt{\frac{2}{\gamma\tilde{\beta}}}
\frac{\partial\eta_{\mathrm m0}}{\partial r}\frac{\partial B_{r0}}
{\partial z}-\sqrt{\frac{2}{\gamma\tilde{\beta}}}\eta_{\mathrm m0} \frac{\partial^2 B_{z1}}{\partial z^2}
\label{vertind1}
\end{equation}
which gives, with $B_{r0}=0$ from Eq.~\ref{azind0}, that
\begin{equation}
\frac{\partial^2 B_{z1}}{\partial z^2}=0.
\label{vertind1b}
\end{equation}
Since we already obtained $B_{z1}=f(r)$ from Eq.~(\ref{divb2}), there is no new constraint here.

\subsubsection*{Order $\epsilon^{2}$:}
\begin{equation}
\begin{aligned}
B_{z0}\frac{\partial\varv_{r1}}{\partial r}+\frac{B_{z0}\varv_{r1}}{r}+\varv_{r1}\frac{\partial B_{z0}}{\partial r}=\\
\sqrt{\frac{2}{\gamma\tilde{\beta}}}\left[\frac{\partial\eta_{\mathrm m0}}{\partial r}\left(\frac{\partial B_{z0}}{\partial r}-\frac{\partial B_{r1}}{\partial z}\right)+\eta_{\mathrm m0}\left(\frac{\partial B_{z0}}{r\partial r}+\frac{\partial^2 B_{z0}}{\partial r^2}\right)\right].
\end{aligned}
\label{vertind2b}
\end{equation}

\subsection*{Energy equation:}
\begin{equation}
\begin{aligned}
\epsilon^4 \tilde{R}^2\tilde{\Omega}^2 n\rho \varv_r\frac{\partial c_s^2}
{\partial r}+\epsilon^3\tilde{R}^2\tilde{\Omega}^2 n
\rho \varv_z\frac{\partial c_s^2}{\partial z}
+\epsilon\rho\tilde{\varv}^2\varv_z \frac{\partial \varv^2/2}{\partial z}\\
+\epsilon^2\rho\tilde{\varv}^2_{\rm A}\varv_r\frac{\partial \varv_{\rm A}}{\partial r}
+\epsilon\rho\tilde{\varv}_{\rm A}^2\varv_z
\frac{\partial \varv_{\rm A}^2}{\partial z}
+\epsilon^2\rho\tilde{\varv}^2\varv_r\frac{\partial \varv/2}{\partial r}\\
+\left[\epsilon^2\rho\tilde{\Omega}^2 \tilde{R}^2\frac{\varv_r}{r^2}+\epsilon^3\rho\tilde{\Omega}^2\tilde{R}^2\frac{\varv_z z}{r^3}\right]
\left[1+\epsilon^2\left(\frac{z}{r} \right)^2\right]^{-3/2}\\
=\frac{\tilde{\varv}^2_{\mathrm A}B_r}{r}
\left(\epsilon^2 \varv_r B_r+\epsilon\Omega rB_\varphi\right)\\
+\tilde{\varv}^2_{\rm A}\frac{\partial}
{\partial r}\left(\epsilon^2 \varv_r B_r^2 +
\epsilon^2 \varv_z B_z B_r +\epsilon\Omega r B_{\varphi} B_r \right) \\
+\tilde{\varv}^2_{\rm A}\frac{\partial}{\partial z}\left( \epsilon \varv_r B_r B_z+
\epsilon \varv_z B_z^2+\Omega r B_{\varphi} B_z\right)
\end{aligned}
\end{equation}

\subsubsection*{Order $\epsilon^{0}$:}
\beqa
0=\tilde{\varv}^2_{\rm A}\frac{\partial}{\partial z}(\Omega_0 r_0 B_{\varphi 0}
B_{z0})\Rightarrow\frac{\partial B_{\varphi 0}}{\partial z}=0,\nonumber
\label{eneq0}
\eeqa

\subsubsection*{Order $\epsilon^{1}$:}
\begin{equation}
\begin{aligned}
\frac{B_{r0}}{2r}+\frac{\partial B_{r0}}
{\partial r}+\frac{\partial B_{z1}}{\partial z}
+\frac{B_{z0}}{B_{\varphi 0}} \frac{\partial B_{\varphi 1}}{\partial z} = 0
\end{aligned}
\label{eneq1}
\end{equation}
Indeed, since $B_{r0}=\partial_zB_{z1}=0$ this gives:
\beq
\frac{\partial B_{\varphi 1}}{\partial z}=0\Rightarrow B_{\varphi 1}=f(r).
\label{bvarphi1}
\eeq

\subsubsection*{Order $\epsilon^2$:}
\beq
\begin{aligned}
B_{\varphi 1}(B_{r0}+B_{r1})+r^{5/2}\frac{\partial}{\partial
r}[\varv_{r2}B_{r0}B_{z0}+\varv_{r1}(B_{r1}B_{z0}\\
+B_{r0}B_{z1})+\frac{1}{r^{1/2}}(B_{z0}B_{\varphi 2}+B_{z1}B_{\varphi 1})+r\Omega_2B_{z0}B_{\varphi 0}]=0,\nonumber
\end{aligned}
\eeq
which becomes, with $B_{r0}=0$:
\beq
\begin{aligned}
B_{\varphi 1}B_{r1}+r^{5/2}\frac{\partial}{\partial r}
[\varv_{r1}B_{r1}B_{z0}+\frac{1}{r^{1/2}}(B_{z0}B_{\varphi 2}+B_{z1}B_{\varphi 1})\\
+r\Omega_2B_{z0}B_{\varphi 0}]=0.
\label{eneq2}
\end{aligned}
\eeq

\section{Solutions and matching functions}

\begin{figure*}
\centering
\includegraphics[width=\columnwidth]{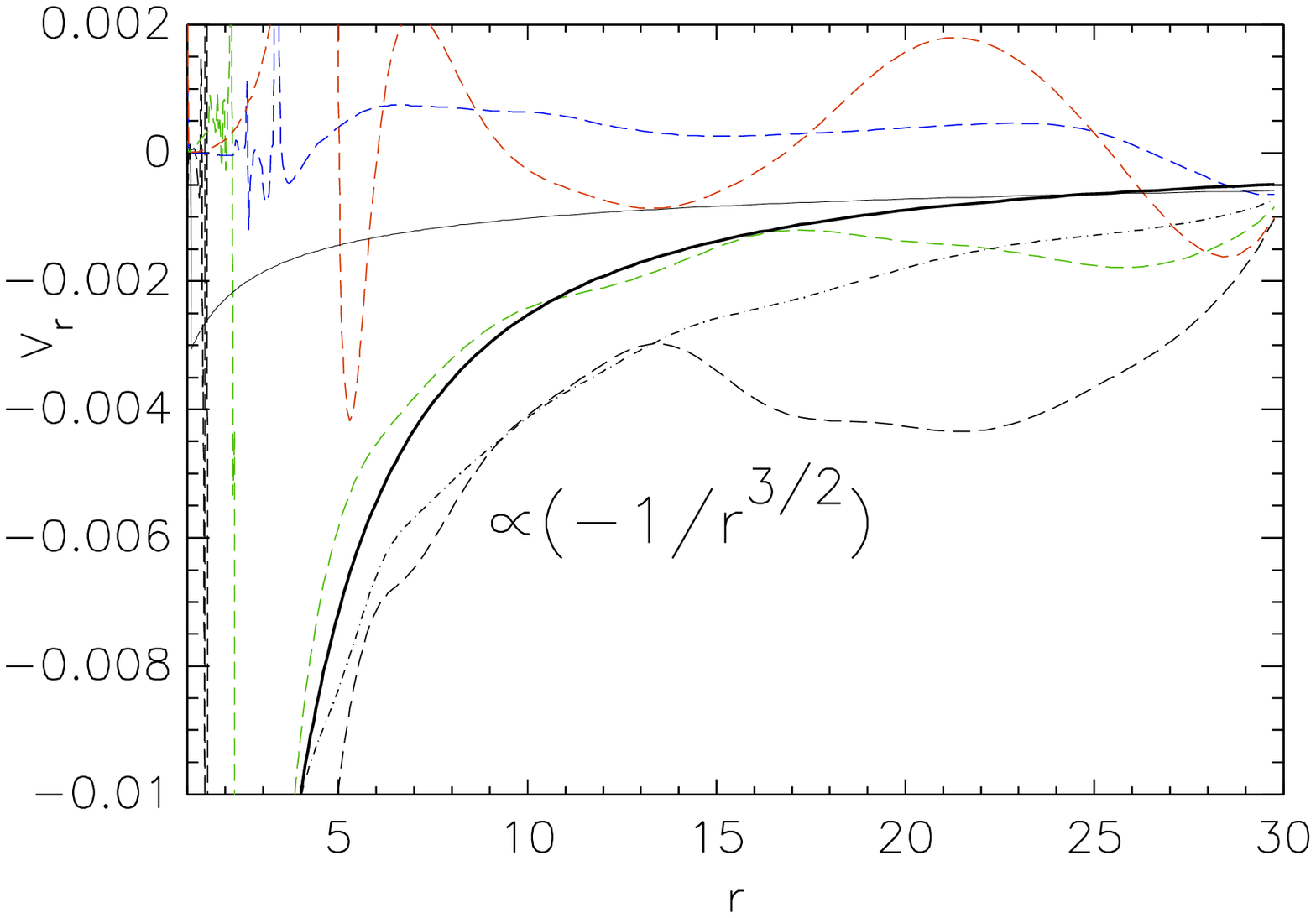}
\includegraphics[width=\columnwidth]{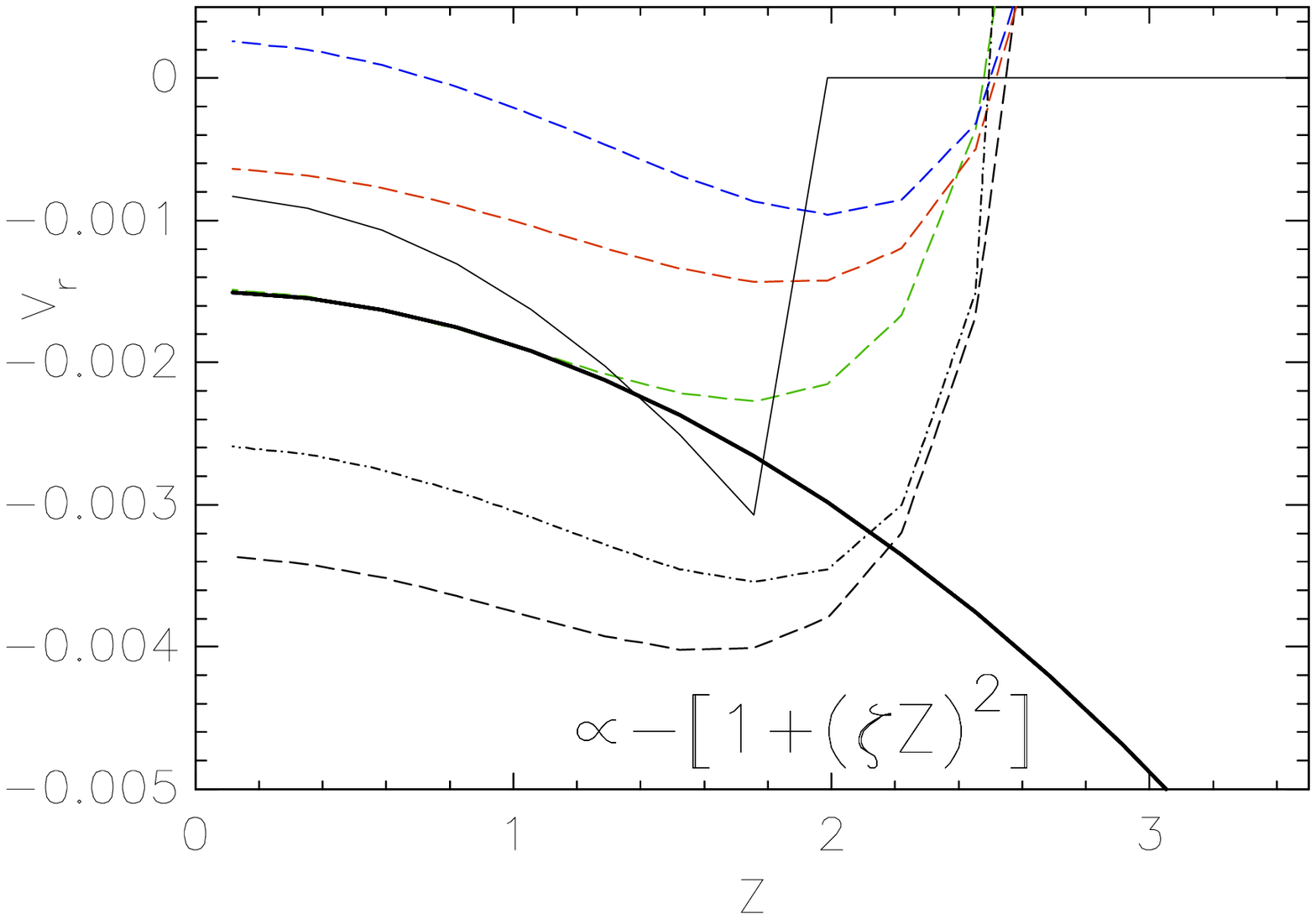}
\includegraphics[width=\columnwidth]{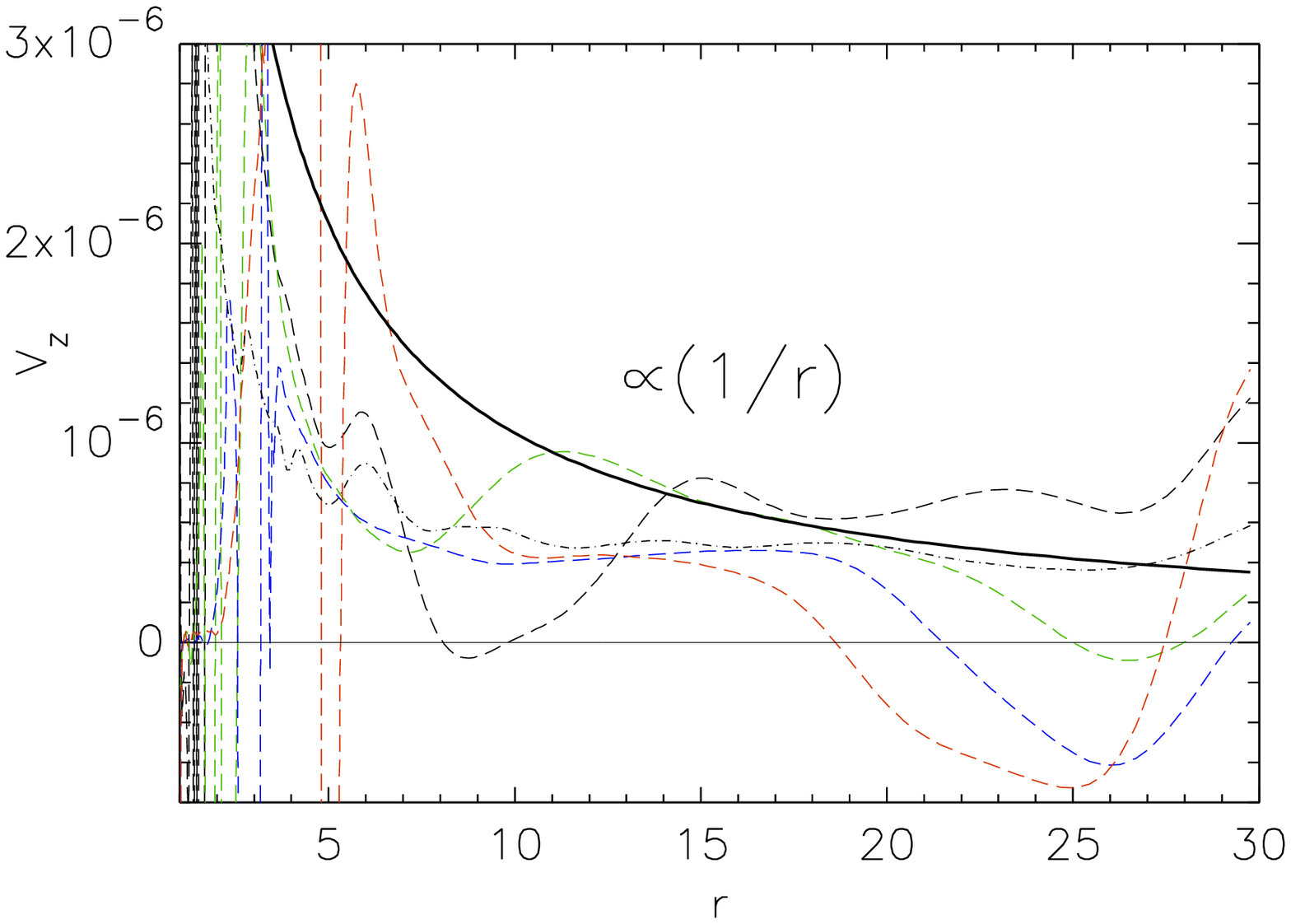}
\includegraphics[width=\columnwidth]{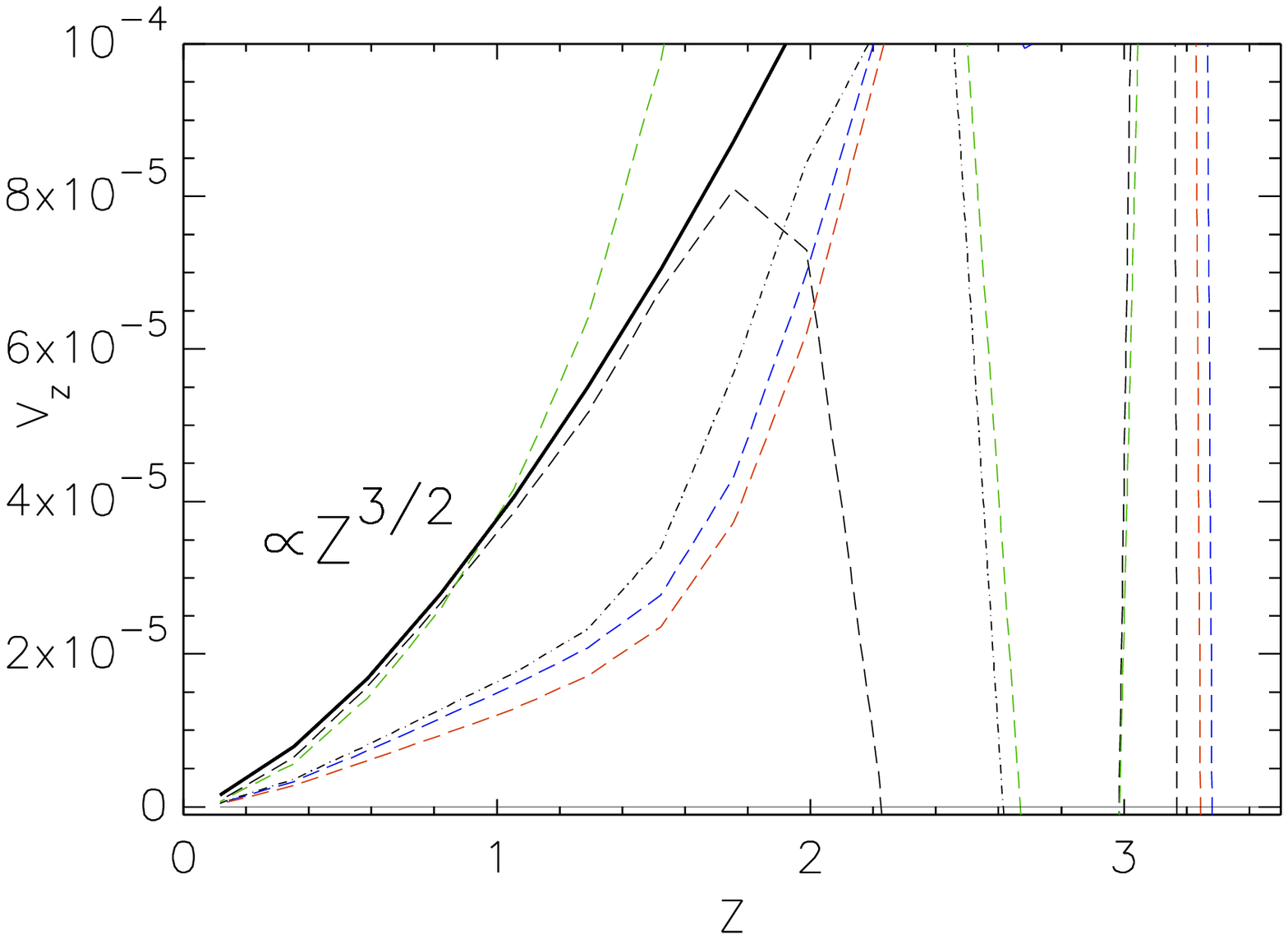} 
\includegraphics[width=\columnwidth]{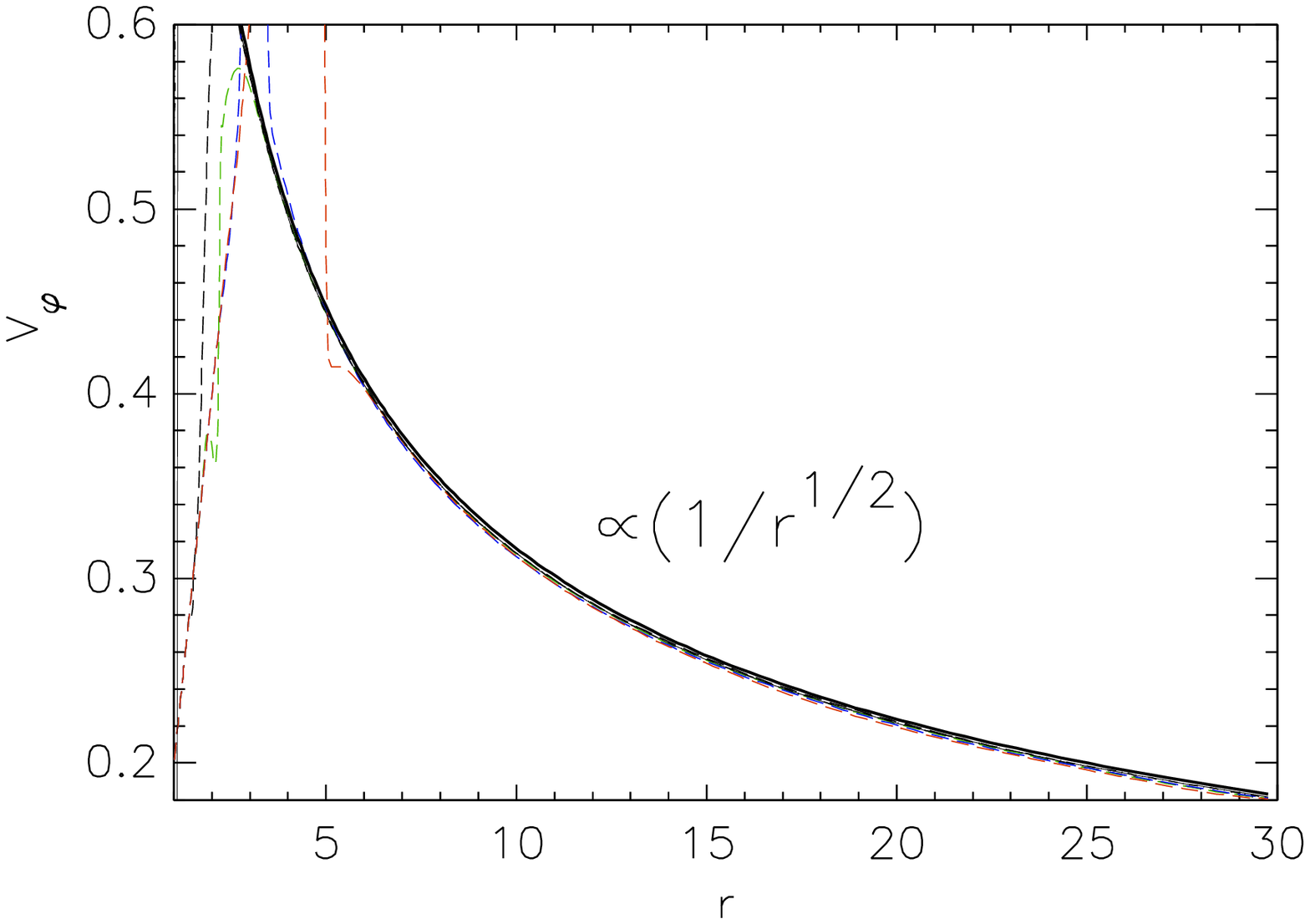}
\includegraphics[width=\columnwidth]{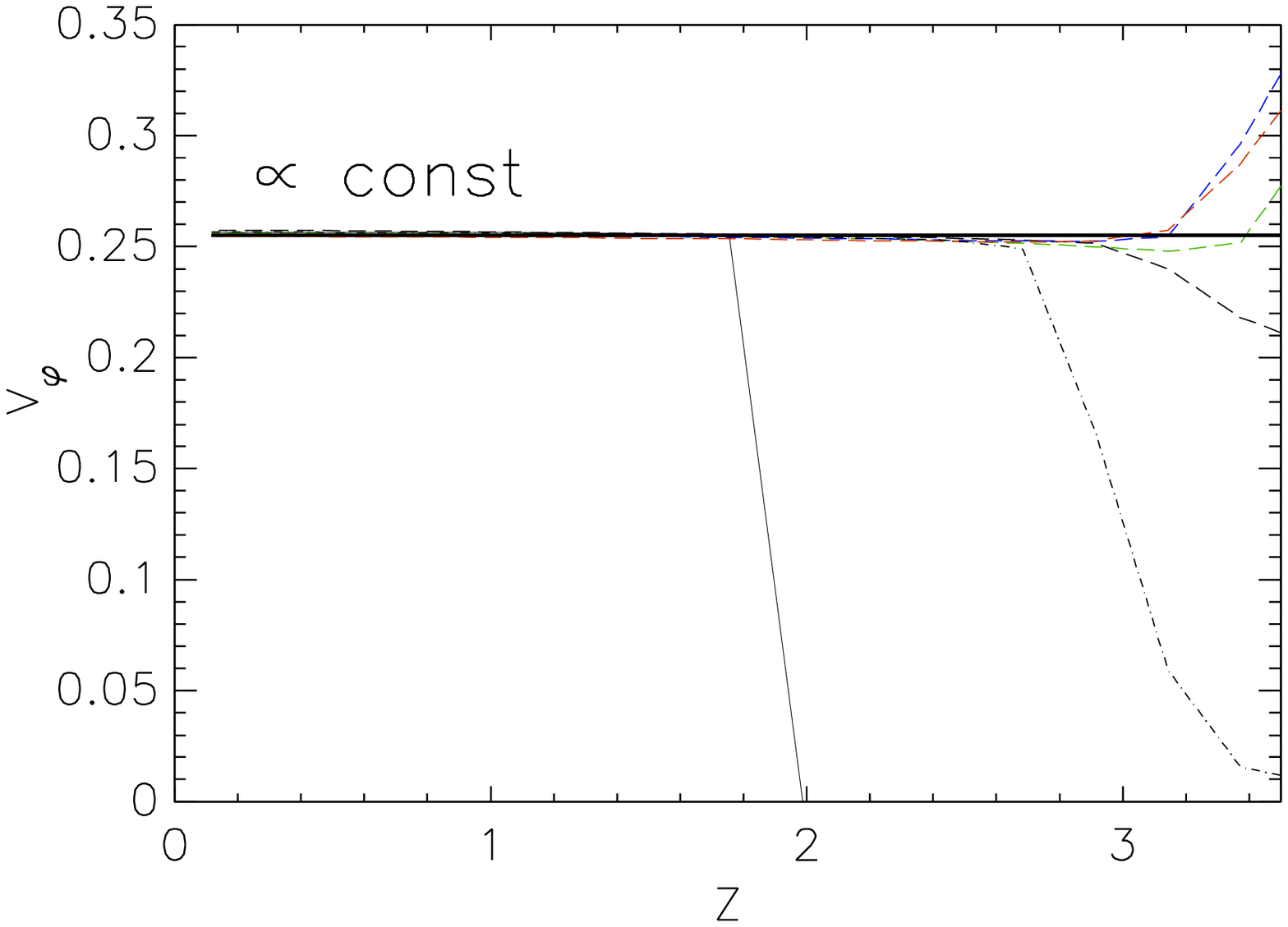} 
\caption{Comparison of the velocity components in the initial set-up
(thin solid line) with the quasi-stationary solutions in the numerical
simulations in the HD (dot-dashed line) and the MHD (long-dashed line)
cases, with $\Omega=0.2\Omega_{\mathrm br}$. The meaning of the lines
is the same as in Fig.~\ref{fig:comparerho1}.}
\label{fig:compare1}
\end{figure*}
\begin{figure*}
\centering
\includegraphics[width=\columnwidth]{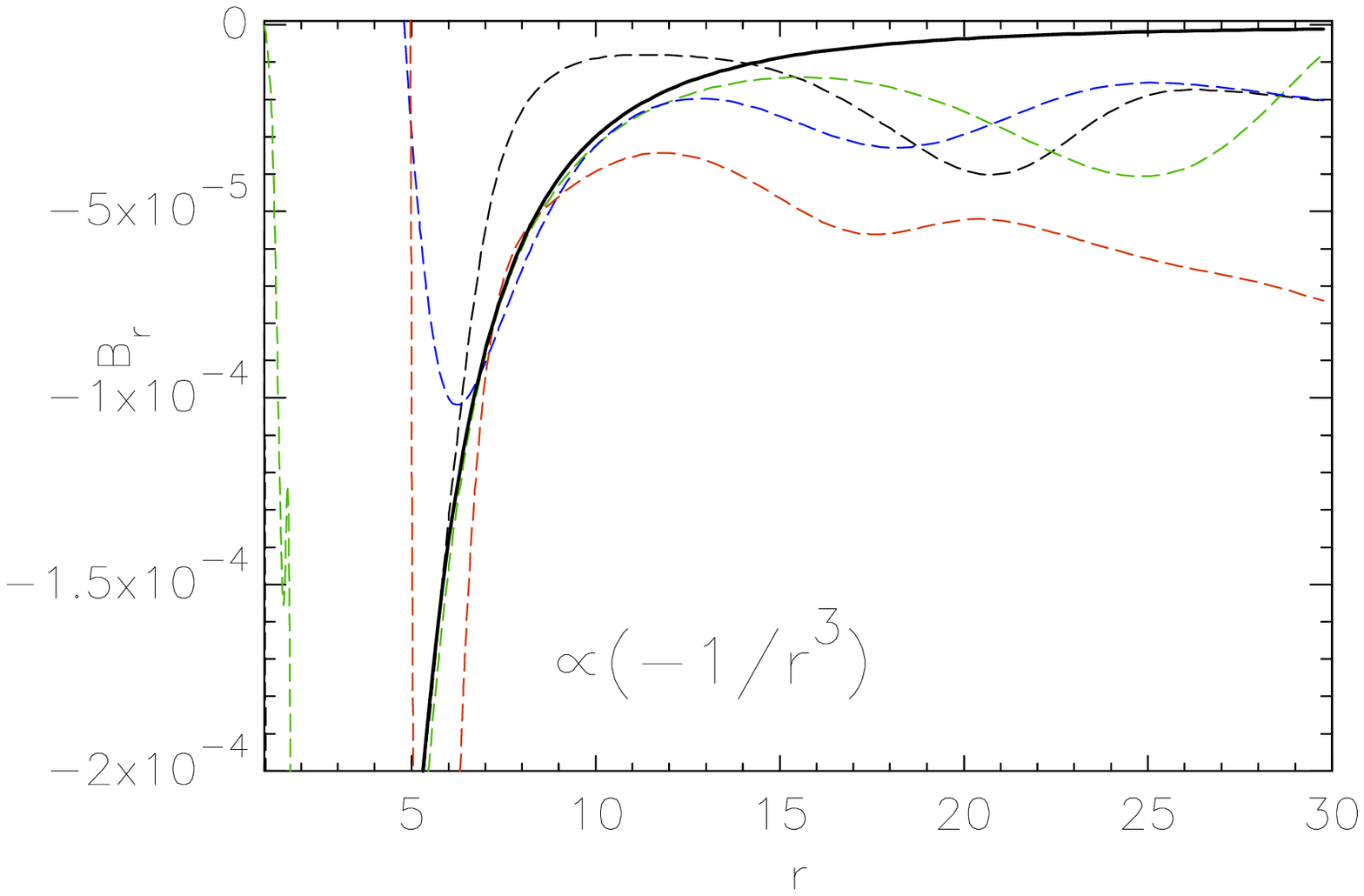}
\includegraphics[width=\columnwidth]{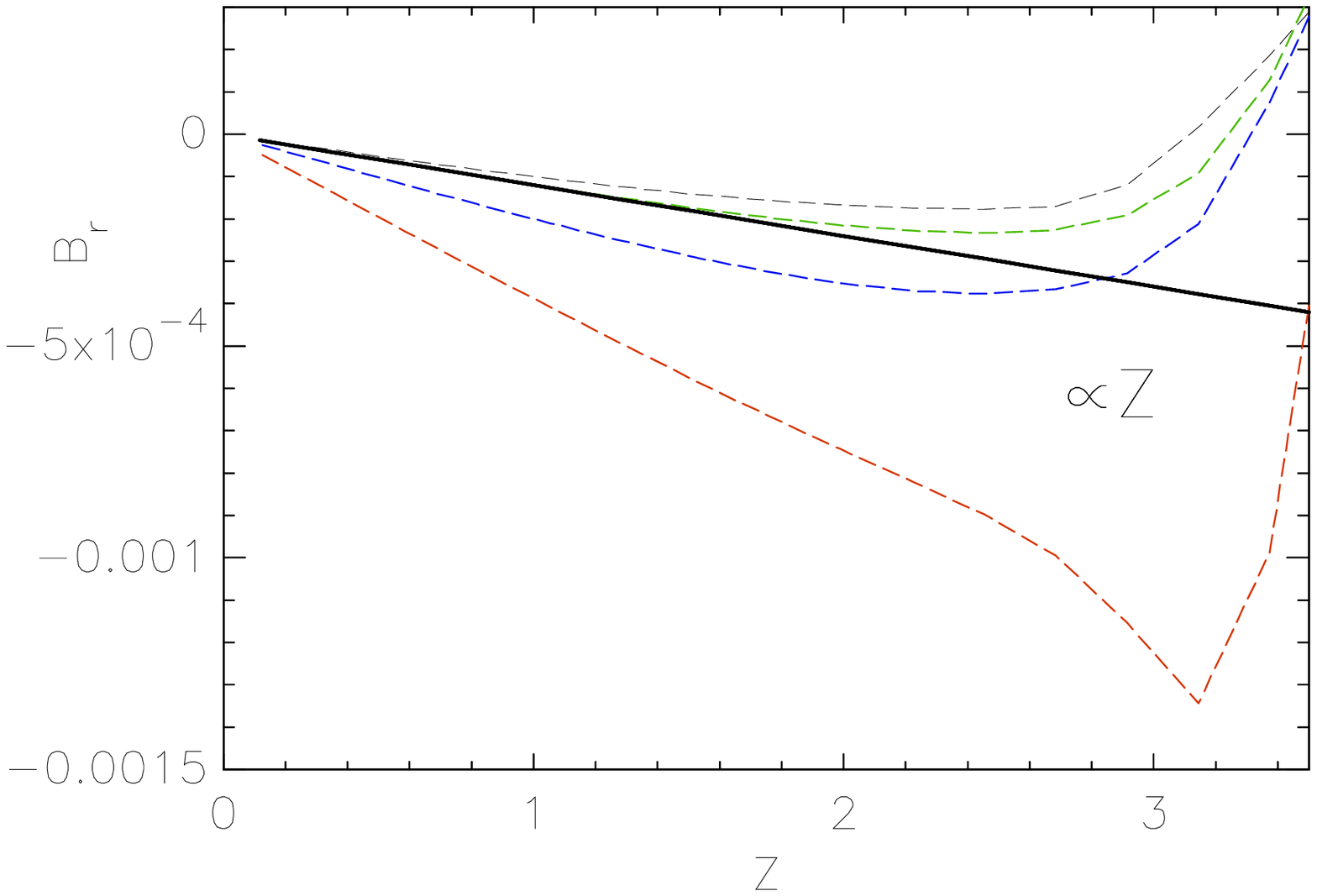}  
\includegraphics[width=\columnwidth]{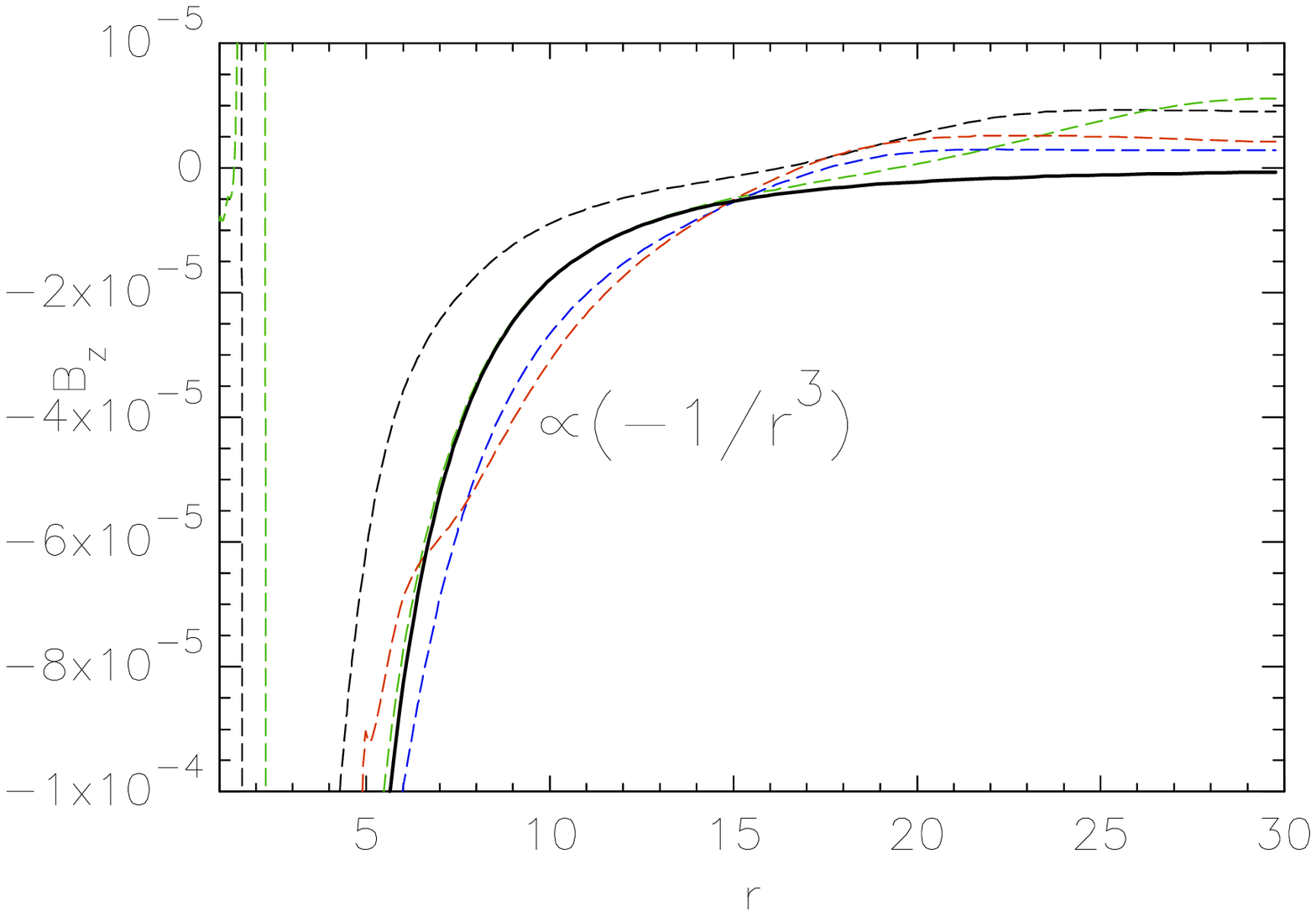}
\includegraphics[width=\columnwidth]{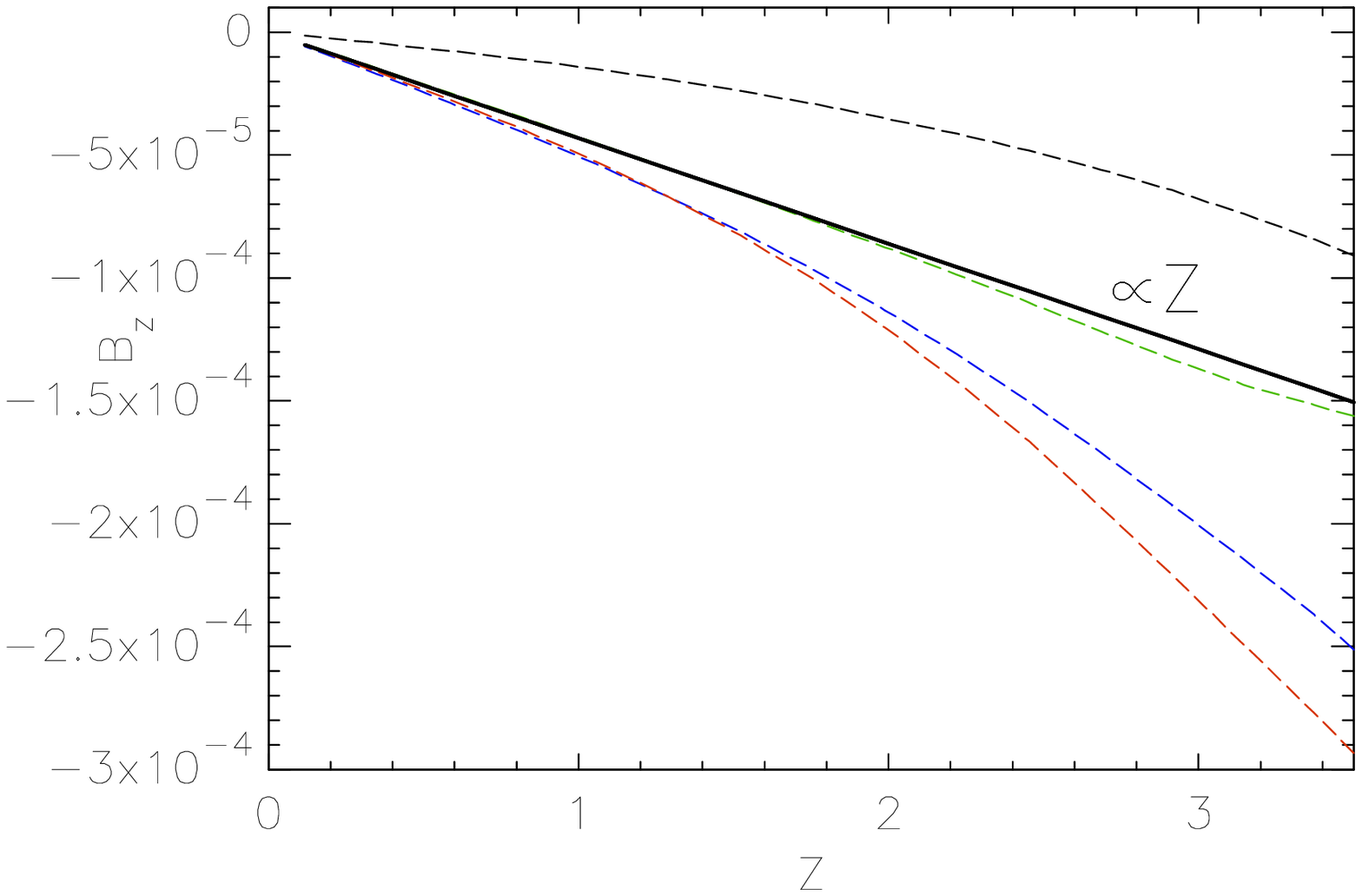}  
\includegraphics[width=\columnwidth]{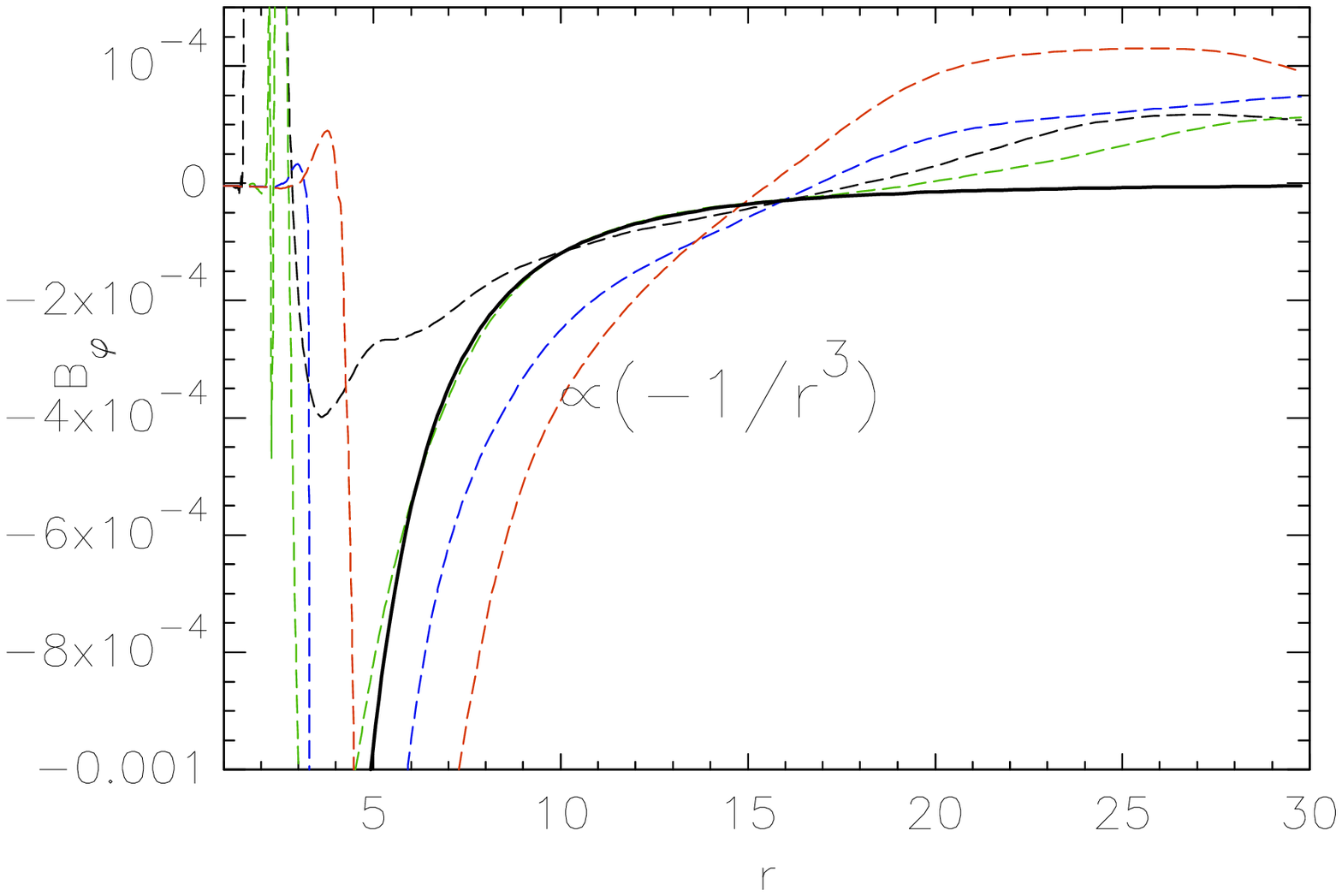}  
\includegraphics[width=\columnwidth]{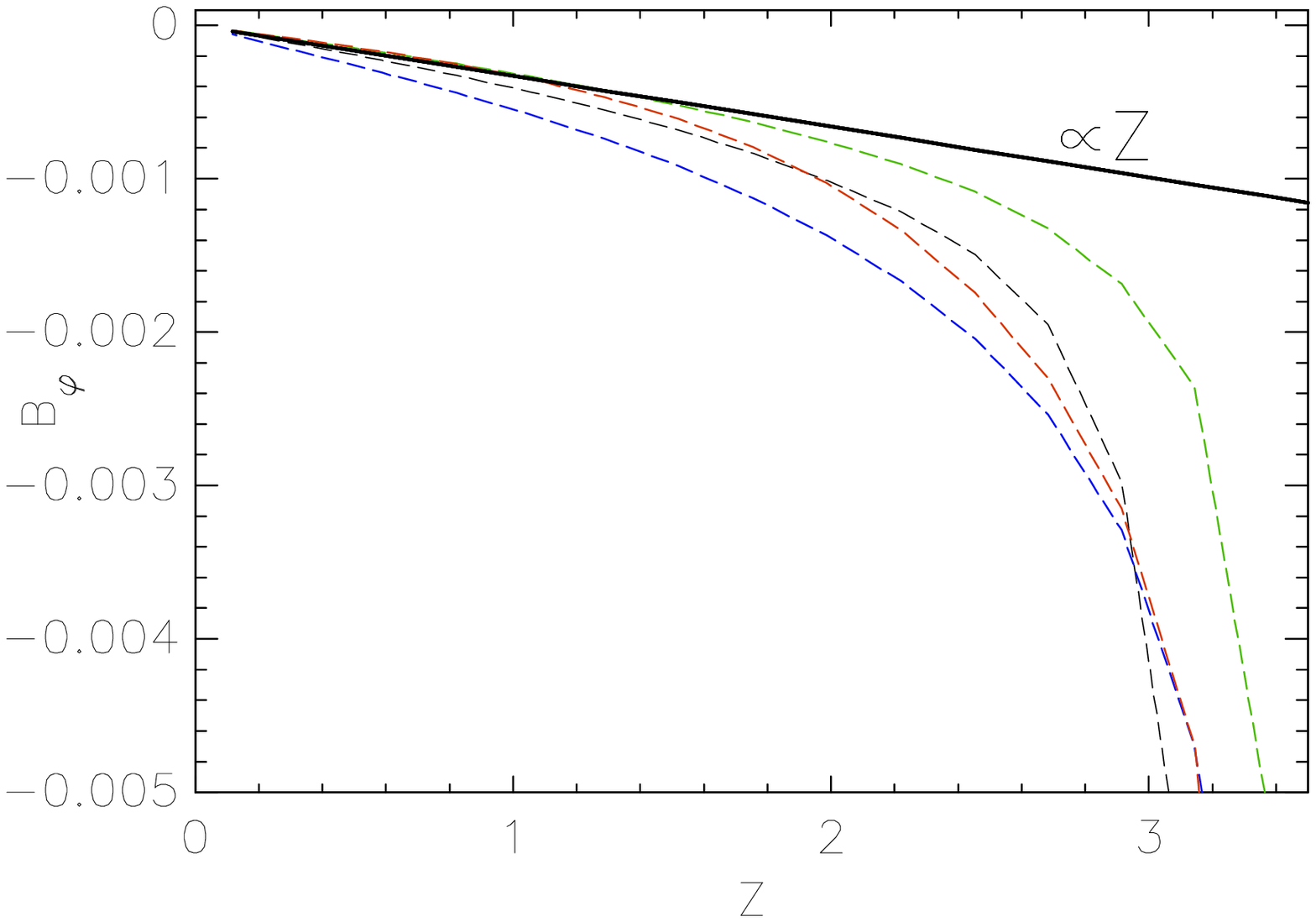}
\caption{The radial, vertical and azimuthal components of the magnetic
field. The meaning of the lines is the same as in
Fig.~\ref{fig:comparerho1}.}
\label{fig:compare2}
\end{figure*}
\begin{figure*}
\centering
\includegraphics[width=\columnwidth]{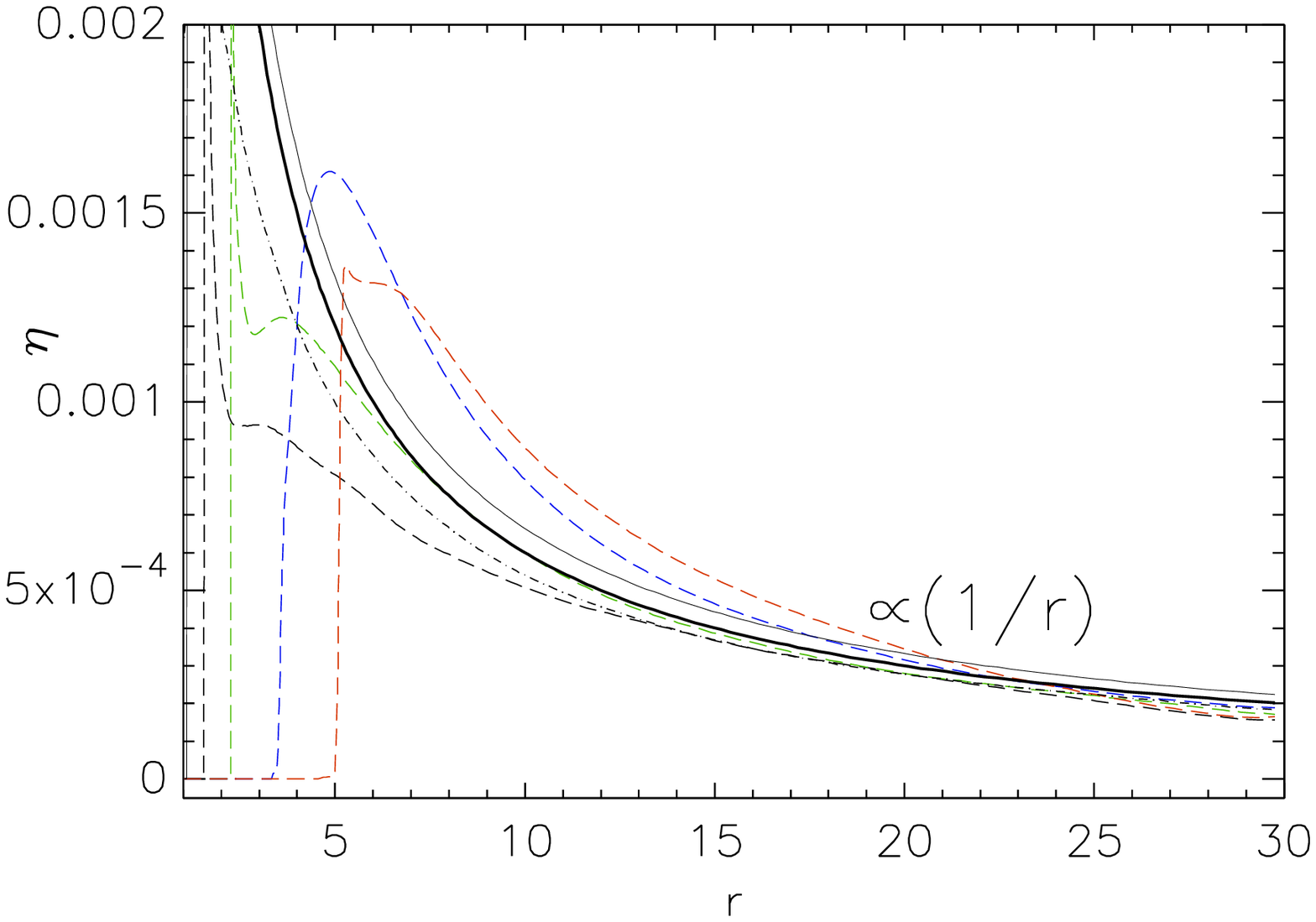}
\includegraphics[width=\columnwidth]{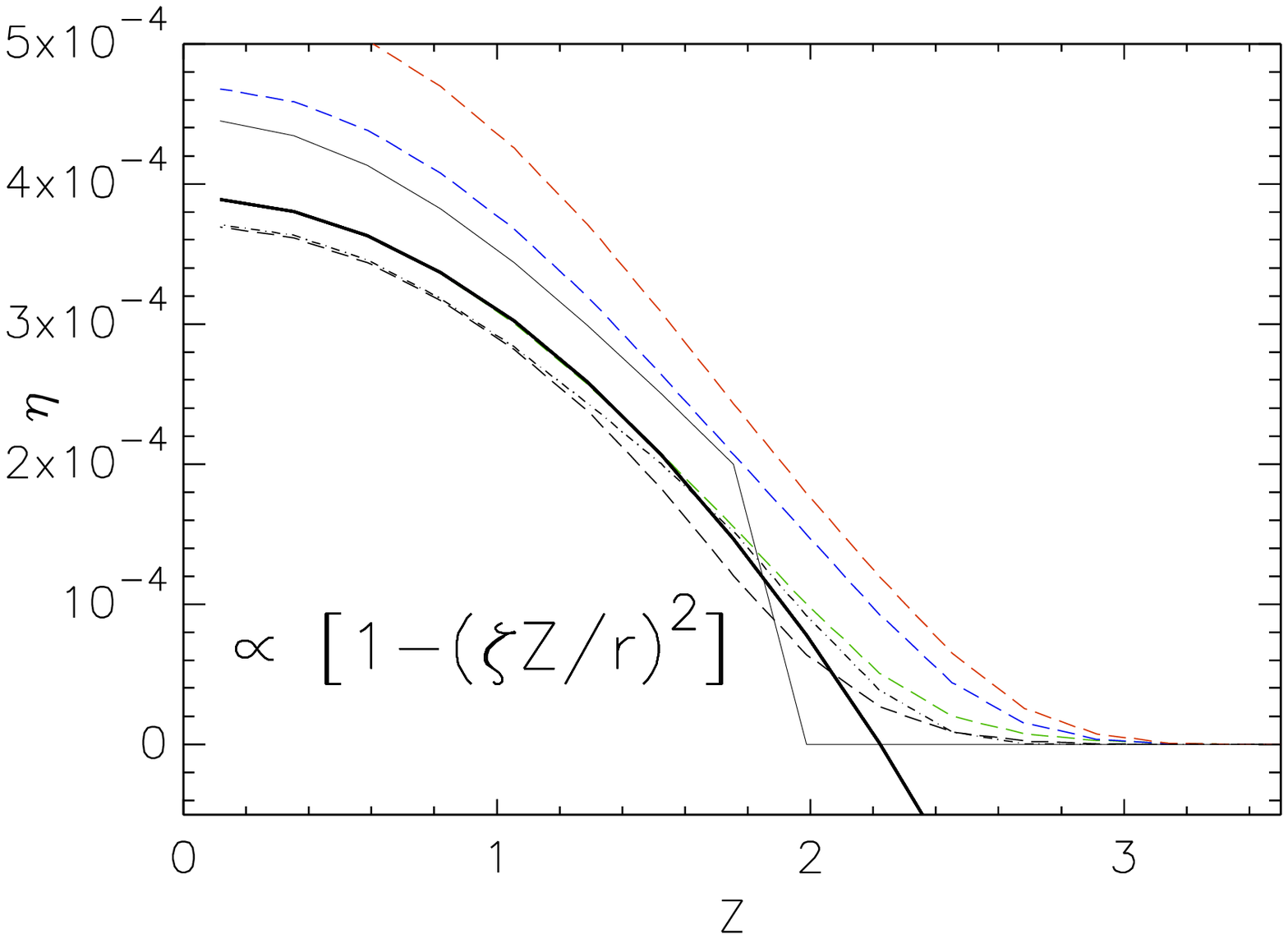}
\includegraphics[width=\columnwidth]{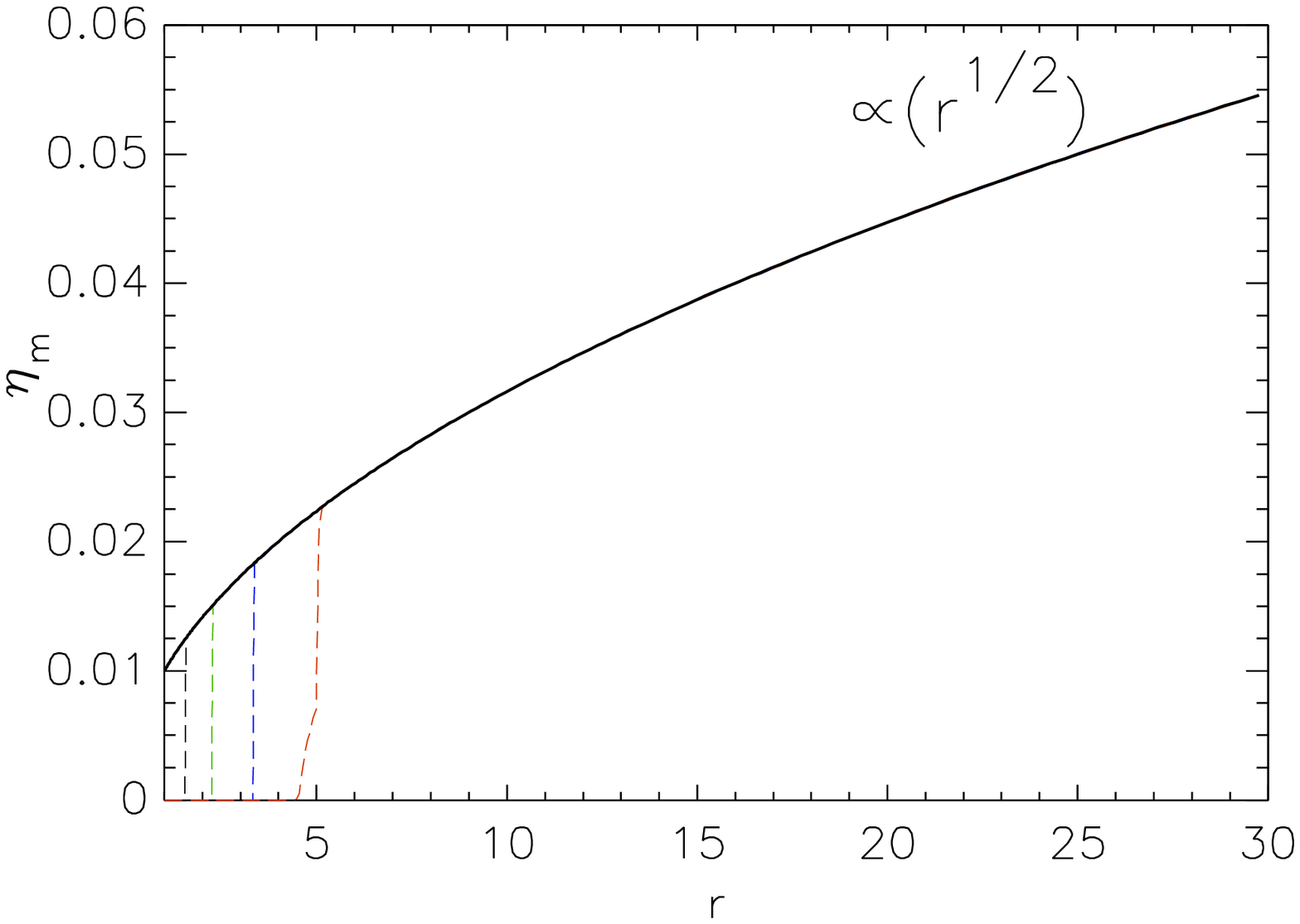} 
\includegraphics[width=\columnwidth]{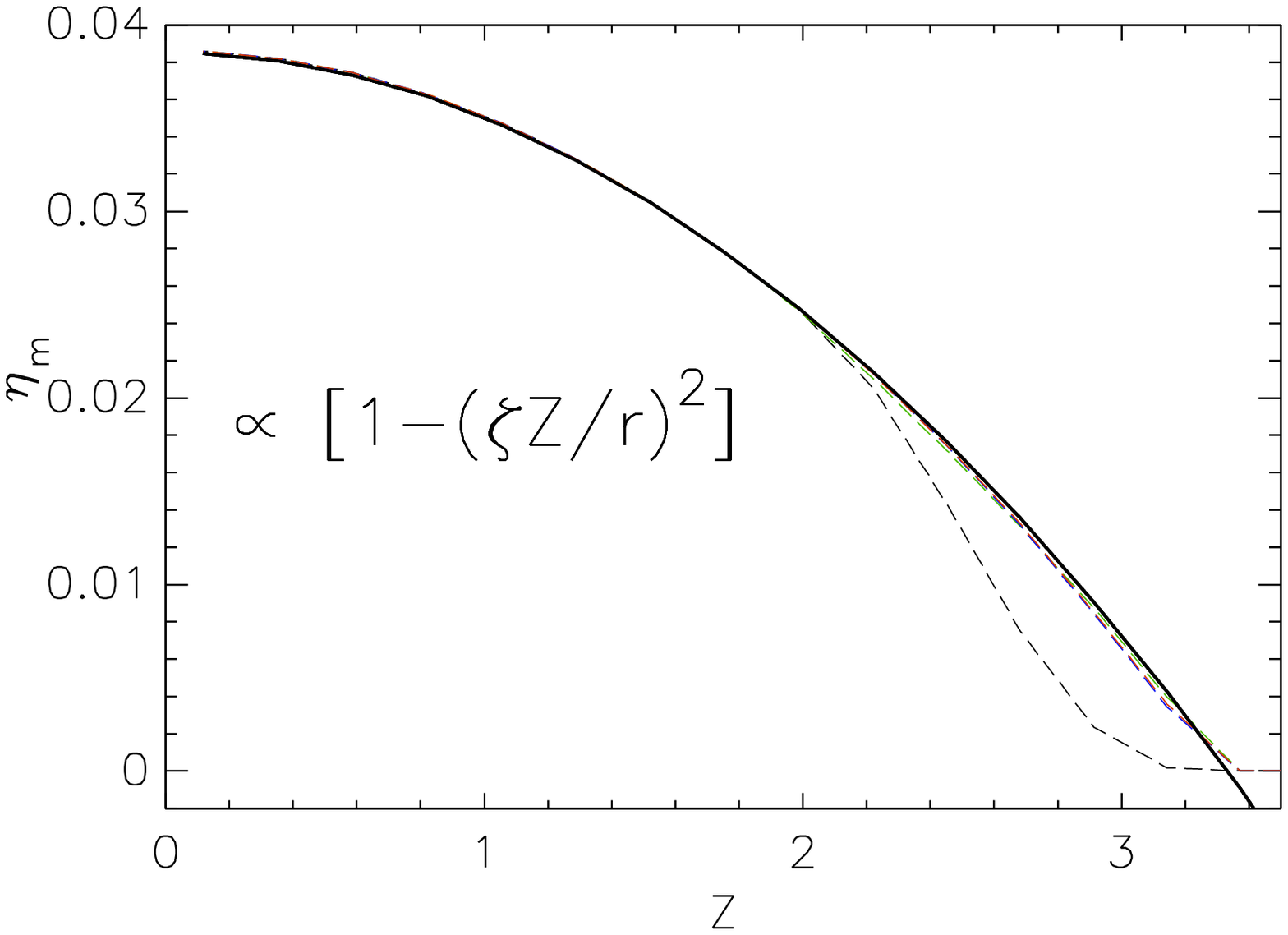}
\caption{Comparison of the results for the anomalous viscosity ($\eta$)
and resistivity ($\eta_m$) coefficients. In the left panels are shown the
radial profiles along the disk equator, and in the right panels the
profiles along the vertical line at r=15$R_\star$. The lines have the
same meaning as in Fig.~\ref{fig:comparerho1}. In the resistivity
profiles in the bottom panels, all the lines overlap along a part of the
profile.}
\label{fig:compareetas}
\end{figure*}
We show here the results in the cases of YSOs with the stellar magnetic 
field of 250, 500, 750 and 1000~G. In all the figures, shown are the
approximate matching curves to the MHD solution in the case with the
stellar field of 500~G.

When there are no oscillations in the solution, matching curves are mostly
inside the 10 per cent error margin. When the oscillations are present, the error
is larger. Functions are chosen to best match the values in the region of
interest in the respective slices, even when it results in a larger error
in the other parts of the approximated line.

In Figs.~\ref{fig:comparerho1} and \ref{fig:compare1}-\ref{fig:compareetas}
are shown the (cylindrical) radius and vertical direction profiles of the
density, velocity, magnetic field components, viscosity and resistivity.
In the radial direction, the values are taken just above the disk midplane
(along the spherical radius line just above $\theta$=$90^\circ$). In the
vertical direction, shown are the slices along lines at half of the disk
length in our simulations ($r$=$15R_\star$).

\end{document}